**Title: Exploiting the geometry of heterogeneous networks: A case study of the Indian stock market.**

Pawanesh Pawanesh[1, #a], Charu Sharma[1, #a] and Niteesh Sahni[1, *#a]

[1]Department of Mathematics, Shiv Nadar Institution of Eminence Deemed to be University, Gautam Buddha Nagar, Uttar Pradesh, India

[#a]Current Address: Department of Mathematics, Shiv Nadar Institution of Eminence Deemed to be University, Gautam Buddha Nagar, Uttar Pradesh, India

[*]Corresponding Author

E-mail: niteesh.sahni@snu.edu.

Phone: +91-9873464028

# Abstract

In this study, we model the Indian stock market as heterogenous scale free network, which is then embedded in a two dimensional hyperbolic space through a machine learning based technique called as coalescent embedding. This allows us to apply the hyperbolic kmeans algorithm on the Poincare disc and the clusters so obtained resemble the original network communities more closely than the clusters obtained via Euclidean kmeans on the basis of well-known measures normalised mutual information and adjusted mutual information. Through this, we are able to clearly distinguish between periods of market stability and volatility by applying non-parametric statistical tests with a significance level of 0.05 to geometric measures namely hyperbolic distance and hyperbolic shortest path distance. After that, we are able to spot significant market change early by leveraging the Bollinger Band analysis on the time series of modularity in the embedded networks of each window. Finally, the radial distance and the Equidistance Angular coordinates help in visualizing the embedded network in the Poincare disc and it is seen that specific market sectors cluster together.

# 1 Introduction

In the recent past, complex networks have emerged as a powerful tool to answer many fundamental questions about the stock market. Several authors have incorporated pairwise correlations between the rate of returns of stocks to model the pairwise interactions and constructed the networks (Jiang et al. 2014; Millington and Niranjan 2017; Sharma and Habib 2019; Guo et al. 2018; Aste et al. 2010; Pozzi et al. 2013; Chi et al. 2010; Kumar and Deo 2012; Tumminello et al. 2005; Nie and Song 2018). Other popular choices of capturing the interaction between pairs of stocks have been the Kendall's tau (Millington and Niranjan 2017), and Mutual information (Chi et al. 2010; Kumar and Deo 2012). A large body of literature exploits the topology of financial networks (Jiang et al. 2014; Millington and Niranjan 2017; Sharma and Habib 2019; Guo et al. 2018; Aste et al. 2010; Pozzi et al. 2013; Chi et al. 2010; Kumar and Deo 2012; Tumminello et al. 2005; Nie and Song 2018) but very few have investigated the geometry of such networks. Only a handful of work with the financial networks embedded in Euclidean space (Nie and Song 2018; Sharma et al. 2017). The above mentioned financial networks are essentially heterogeneous scale free networks. It is interesting to note that heterogeneous scale free networks emerge naturally in a variety of domains like Infrastructure, Communication, Biology, and Social media. In the networks modelling key scenarios in these areas, assessment of the health of the system plays a fundamental role in predicting breakdowns whether through outside attacks or a system failure. One way to address such problems is to solve an optimal percolation problem which is generally NP hard. It has been shown in (Artime et al. 2024) that an approximate solution to this optimal percolation problem is possible by considering a topology based measure called as Collective Influence (CI) of each node in the network. For precise details and simulation results in a variety of scenarios, the reader may refer to (Artime et al. 2024). However there are number of scenarios in Epidemiology, Medicine, Engineering etc. where it is imperative to study the network dynamics. To this end signal propagation in heterogeneous networks is analysed using modelling and data driven tools including Complex networks, Statistics, Machine learning, Time series etc. For a comprehensive survey of these tools the reader may refer to (Ji et al. 2023).

Hyperbolic spaces have been shown to provide a more accurate representation of many topological-world networks than Euclidean spaces (Keller-Ressel and Nargang 2021; García-Pérez et al. 2016; Muscoloni et al. 2017; Cacciola et al. 2017; Zhou and Sharpee 2021).

In fact, very recently Keller-Ressel and Nargang (2021) studied the latent geometry corresponding to the networks of European banks for the various years by embedding the underlying networks of banks in the Poincaré disc. The vertices in the network were the banks and the weights associated with each edge was the liquidity-weighted portfolio overlap of one bank with the other bank. It measures the impact of sudden liquidation of portfolio of one bank on the other bank and vice versa. After embedding, each bank is assigned a radial coordinate $r$ and an angular coordinate $\theta$. Statistical methods are employed to establish associations between the radial coordinates and the systemic importance of the banks (known from the FSB reports) and also between the angular coordinates and regional sub sectors.

In reference (García-Pérez et al. 2016), authors have an empirical evidence that the world trade network obey the power-law, that is they are heterogeneous. Based on this evidence, they drew inspiration from (Krioukov et al. 2010) and consequently adopted the connectivity probability while following the steps outlined in the referenced paper's proof to derive the radial and angular coordinates for each vertices in the Poincaré disc.

A body of recent literature including (Chen and Hao 2020; Chen et al. 2024; Sivadasan et al. 2024) is devoted to building deep learning models based on Artificial Neural Networks and Recurrent Neural Networks for predicting stock market risks and prices. Our method is based on the idea of embedding a heterogeneous network in a 2D Hyperbolic space through a machine learning technique called Coalescent embedding (Muscoloni et al. 2017) developed recently.

An interesting application of coalescent embedding in medicine is worth pointing out wherein, normal patients are distinguished from the ones ailing with Parkinson's disease on the basis of the anatomical changes occurring in the human brain over time (Cacciola et al. 2017). The edges are defined purely on the basis of anatomy rather than defined on the basis of a mathematical rule unlike the application considered in this paper. It has been established that the regions of the brain (vertices) belonging to different lobes of the brain are clearly separated in the Poincaré disc.

As explained in the Preliminaries section, a Poincaré disc is a natural framework to embed networks exhibiting a power law. Based on the community analysis and cluster analysis of the embedded networks we are able to segregate stocks of different sectors and discover important patterns within a sector. Refer to section 5 for complete details. Embedding allows us to compute geometric measures like hyperbolic distance (HD) and hyperbolic shortest path distance (HSD) which have been shown to segregate the periods of volatility and stability of the markets via non-parametric statistical tests. For complete detail refer to section 5.2.

The standard method of finding communities in a network is to use the classical algorithm of Newman (Newman 2004). These communities will be referred to as topological communities throughout this paper and the family of topological communities will be denoted by $\mathcal{C}^{top}$. In (Nie and Song 2018) the authors embed a network into a Euclidean space via a nonlinear dimensionality reduction algorithm such as Isomap (ISO) (Tenenbaum et al. 2000) and subsequently use k-means algorithm to compute clusters denoted as $\mathcal{C}^{Euc}$. Finally, the similarity between $\mathcal{C}^{top}$ and $\mathcal{C}^{Euc}$ is adjusted on the basis of the Normalised mutual Information (NMI) (Danon et al. 2005; Danon et al. 2006). In the present paper, we apply the hyperbolic version of k-means algorithm (Nickel and Kiela 2017) to the network embedded in the hyperbolic disc to obtain the clusters and call the clusters thus obtained as the hyperbolic clusters and denote the family of hyperbolic clusters by $\mathcal{C}^{hyp}$. Further, we compare the clusters on the basis of the Normalised mutual Information (NMI) as well as adjusted mutual Information (AMI) (Vinh et al. 2009). It has been observed that $NMI(\mathcal{C}^{top}, \mathcal{C}^{hyp})$ (as well as $AMI(\mathcal{C}^{top}, \mathcal{C}^{hyp})$) is far higher than the $NMI(\mathcal{C}^{top}, \mathcal{C}^{Euc})$ (as well as $AMI(\mathcal{C}^{top}, \mathcal{C}^{hyp})$). Thus, the topological community structure matches the hyperbolic clusters far more significantly than the ones obtained upon Euclidean embedding.

We begin our analysis by modelling the stock market data as a complex network by considering pairwise correlations between stocks. It is worth noting that the full correlation-based network has a homogeneous degree distribution and hence does not obey the power law. Thus, we need to work with certain subgraphs of the full network. The Minimum Spanning Tree (MST), the Planar Maximally Filtered Graph (PMFG), and the PMFG-based threshold network (PTN) have been popular network models studied in a huge body of research over the years (Jiang et al. 2014; Millington and Niranjan 2017; Sharma and Habib 2019; Guo et al. 2018; Aste et al. 2010; Pozzi et al. 2013; Chi et al. 2010; Kumar and Deo 2012; Tumminello et al. 2005; Nie and Song 2018). We shall also resort to these structures as our starting point.

As mentioned in Preliminary Section, any network for which the degree distribution of the vertices follow the power law can be embedded in the hyperbolic disc. Thus we carry out statistical computations to verify the validity of the power law for each of our networks – MST, PMFG, and PTN.

Angular separation index (ASI) (Muscoloni and Cannistraci 2019) measures how well the clusters corresponding to the communities are separated in the hyperbolic space. The ASI score is constructed on the basis of counting the number of vertices of other communities that lie between each pair of successive vertices of a given community. Random shuffling of angular coordinates is performed a large number of times to calculate an empirical distribution of the ASI and $p$-values are computed to judge the quality of the observed ASI.

The remaining part of the paper is structured into 5 sections. In section 2, we give a description of the data used in our analysis. In Section 3, we present a preliminary definitions and methods utilized in our study. In section 4, we give the methodology adopted in the various analysis undertaken. Section 5 provides a detailed discussion of results. Lastly, in the concluding section, we summarize our key observations and future implication of the work.

# 2 Data description

Our data comprises of daily closing prices of all 500 stocks listed on the CNX500 index of the National Stock Exchange (NSE) India from January 01, 2017, till December 31, 2021. This accounts for 1236 working days. We removed the records of 117 stocks on the account of missing data of prices. This leaves us with the complete data of 383 stocks. Next, we calculate the daily logarithmic returns $r_i(t) = \log\big(p_i(t+1)\big) - \log\big(p_i(t)\big)$. Here $p_i(t)$ stands for the closing price of the $i$-th stock at day $t$. While doing this we remove the data corresponding to non-successive days (like Friday & Monday are separated by more than 1 day because of holidays in between thus we ignore the log return for Monday and similarly there were other exceptions also because of bank holidays). This leaves us with 939 values of logarithmic returns corresponding to the above 383 stocks.

# 3 Preliminaries

The two-dimensional hyperbolic space $\mathbb{H}^2$ (or the Poincaré disc) is represented by the interior of the Euclidean disc of unit radius:

$$\mathbb{H}^2 = \{(x, y) \in \mathbb{R}^2; x^2 + y^2 < 1\}. \tag{1}$$

The hyperbolic distance between the two points $x = (x_1, x_2)$ and $y = (y_1, y_2)$ in $\mathbb{H}^2$ is given by

$$d(x, y) = \cosh^{-1}\left(1 + \frac{2\|x - y\|^2}{(1 - \|x\|^2)(1 - \|y\|^2)}\right). \tag{2}$$

For further details regarding equation 1 and equation 2 refer to (Loustau 2020).
Alternatively, the hyperbolic distance $x_{ij}$ between two points in the Poincaré disc with polar coordinates $(r, \theta)$ and $(r', \theta')$ is given by

$$x_{ij} = \frac{1}{\zeta}\cosh^{-1}(\cosh(\zeta r)\cosh(\zeta r') - \sinh(\zeta r)\sinh(\zeta r')\, cos\Delta\theta). \tag{3}$$

Where $\zeta = \sqrt{-K}$, $K$ is curvature of the hyperbolic space and $\Delta\theta = \pi - |\pi - |\theta - \theta'||$ is the angular distance between points. The hyperbolic distance equation 3 has the following well known approximation:

$$x_{ij} \approx r + r' + \frac{2}{\zeta}\ln\left(\frac{\Delta\theta}{2}\right). \tag{4}$$

For, further details regarding equation 3 and equation 4 refer to (Krioukov et al. 2010).
The boundary $\partial\mathbb{H}^2$ of the disc is given by the circle

$$\mathbb{S}^1 = \{(x, y) \in \mathbb{R}^2; x^2 + y^2 = 1\}.$$

The area and the circumference of the hyperbolic disc $\mathbb{H}^2$ can be given by

$$L(r) = 2\pi \sinh(\zeta r) \propto e^{\zeta r}.$$

$$A(r) = 2\pi(\cosh(\zeta r) - 1) \propto e^{\zeta r}.$$

For the details regarding $\mathbb{S}^1$, $L(r)$ and $A(r)$ refer to (Krioukov et al. 2010)
Hyperbolic spaces are a natural framework for embedding scale free networks which we now explain.
Let the degree distribution of the vertices of the scale free network follow the power law distribution $P(k) \propto k^{-\gamma}$. In order to embed this network in a Poincaré disc, we start by uniformly distributing the $N$ vertices on a circle of radius $R = \frac{N}{2\pi}$ and assign an angular coordinate $\theta$ to each vertex. Let $\kappa$ stand for the expected degree of each vertex in the network. Now corresponding to each vertex we have a pair $(\kappa, \theta)$. Treating $\kappa$ as a hidden variable, we can write

$$P(k) = \int_{\kappa_0}^{\infty} g(k|\kappa)\rho(\kappa)d\kappa. \tag{5}$$

Here $g(k|\kappa) = \frac{e^{-\bar{k}(\kappa)}\left(\bar{k}(\kappa)\right)^k}{k!}$. For, further details regarding the expression $g(k|\kappa)$ and the equation 5 refer to (Boguná and Pastor-Satorras 2003).

The probability that two vertices $(\kappa, \theta)$ and $(\kappa', \theta')$ are joined by an edge is $\tilde{p}(\chi)$, where $\chi$ is chosen in (Krioukov et al. 2010) as $\chi = \frac{d}{\mu\kappa\kappa'}$, where d is the arc length and the parameter $\mu > 0$. The average degree of vertices with expected degree $\kappa$ is given by

$$\bar{k}(\kappa) = \frac{N}{2\pi}\int_{\kappa_0}^{\infty}\int_0^{2\pi} \rho(\kappa')\tilde{p}\left(\frac{d}{\mu\kappa\kappa'}\right)d\kappa' d\theta'$$

and further it turns out that $\bar{k}(\kappa) = \kappa$. For the details regarding the expression $\bar{k}(\kappa)$ reader refer to (Boguná and Pastor-Satorras 2003).
Now, substituting $\rho(\kappa) = \kappa_0^{\gamma-1}(\gamma-1)\kappa^{-\gamma}$, where $\gamma > 2$ and $\kappa \geq \kappa_0$, and the expression for $g(k|\kappa)$ in equation 5 we get

$$\begin{aligned} P(k) &= \int_{\kappa_0}^{\infty} \frac{e^{-\kappa}(\kappa)^k}{k!}\kappa_0^{\gamma-1}(\gamma-1)\kappa^{-\gamma}d\kappa \\ &= \frac{\kappa_0^{\gamma-1}(\gamma-1)}{k!}\int_{\kappa_0}^{\infty} e^{-\kappa}(\kappa)^{k-\gamma+1-1}d\kappa \\ &= \frac{\kappa_0^{\gamma-1}(\gamma-1)}{k!}\Gamma(k-\gamma+1, \kappa_0) \end{aligned}$$

Thus

$$P(k) \propto k^{-\gamma}.$$

Thus every scale free network of a given distribution can be generated by considering the distribution of the expected degree of each vertex.
We now map the point $(\kappa, \theta)$ to a point $(r, \theta)$ in the disc via the transformation:

$$r = R - \frac{2}{\zeta}\log\left(\frac{\kappa}{\kappa_0}\right) \tag{6}$$

(Note that $0 < r < R = 2\ln(\frac{N}{\mu\pi\kappa_0^2})$)
As a consequence of the above transformation we also have

$$\rho(\kappa) = \rho\left(\kappa_0 e^{\frac{(R-r)\zeta}{2}}\right) = \kappa_0^{-1}(\gamma-1)e^{\frac{\gamma\zeta(r-R)}{2}} \propto e^{\frac{\gamma\zeta r}{2}} \tag{7}$$

and $\chi = e^{\frac{\zeta(x-R)}{2}}$, where $x$ is the hyperbolic distance between two vertices. The vertices $(r, \theta)$ and $(r', \theta')$ are connected by an edge according to the connection probability $\tilde{p}(\chi) = H(1 -$ $\chi$). Also the role of the curvature of the hyperbolic space is played by $\frac{\gamma\zeta}{2}$. For further details, regarding the equation 6 and 7 refer to (Krioukov et al. 2010).
On the other hand, a striking consequence of hyperbolic geometry is that $N$ points distributed over the Poincaré disc can be thought of as vertices of a network satisfying the power law. We explain this briefly as under.
Consider Poincaré disc with radius $R$. Assume that the angular coordinates $\theta_i$ of the $N$ points $(r_i, \theta_i)$ are from the uniform distribution $\rho(\theta) = \frac{1}{2\pi}$ on the interval $[0,2\pi]$. The radial density is given by $\rho(r) = \frac{L(r)}{A(R)} \approx e^{r-R}, r \in [0, R]$. These $N$ points constitute the vertex set $\{1,2, \ldots, N\}$ of the graph. Next, there will be an edge between vertices $i$ and $j$ if $x_{ij} \le R$. In other words, the connection probability function takes the form $p(x) = H(R - x)$, where $H(t)$ stands for the heaviside step function. It turns out that $P(k) \propto k^{-3}$. For details the reader refer to (Krioukov et al. 2010).

In this paper we shall work with embeddings of two kinds of sub-graphs: the Minimum spanning tree (MST) and the Planar maximally filtered graph (PMFG).
In MST we retain the most significant edges by defining the edge weight using the distance $\boldsymbol{d_{ij}} = \sqrt{\boldsymbol{2(1-\rho_{ij})}}$, where $\boldsymbol{\rho_{ij}}$ is the Pearson correlation between the log returns of the stock $\boldsymbol{i}$ and stock $\boldsymbol{j}$. Next, Prim's algorithm (Prim 1957) is employed to find a spanning tree with minimum sum of the total edge weights. The importance of using $\boldsymbol{d_{ij}}$ as a weight has been highlighted in (Jiang et al. 2014; Millington and Niranjan 2017; Sharma and Habib 2019; Guo et al. 2018; Aste et al. 2010; Chi et al. 2010; Kumar and Deo 2012; Tumminello et al. 2005; Nie and Song 2018).

The PMFG subgraph of a graph $\boldsymbol{G} = (\boldsymbol{V}, \boldsymbol{E})$ with adjacency matrix $\boldsymbol{A} = [\boldsymbol{a_{ij}}]_{\boldsymbol{n \times n}}$ is extracted as follows:

---

Algorithm 1 for PMFG

---

```
Start
    Extract the upper triangular part of matrix A = [a_ij] and form a list of elements a_ij
    (this is the weight associated with the edge e_ij) and arrange them into a vector S

    Sort the list S in descending order to get list S'
    Initialize an empty graph PMFG
    Add the first six elements from list S' to PMFG along with their corresponding edges
    While the number of edges in PMFG < 3n − 6 (Euler 1958) do
        Take the next element from list S'
        Add this element and its corresponding edge to PMFG temporarily
        If PMFG is planar (using Boyer-Myrvold algorithm (Boyer and Myrvold 2006))
           Retain the added edge
        Else
           Discard the added edge
      End
   End
End
```

---

When this process ends, we are left with a maximally filtered planar graph. It is worth pointing out that the Boyer-Myrvold algorithm implements two steps at its core: first, it creates a straight-line drawing of a planar graph by mapping all $\boldsymbol{n}$ vertices to integer coordinates using an algorithm by Chrobak and Payne (1995). Second, it verifies that the result produced in the first to have a straight-line representation or not. This step uses the Fary's theorem (Fáry 1948).

A brief overview of the key procedures and statistical measures used throughout our analysis are presented in sections 3.1-3.5.

## 3.1 Validity of Power-law in networks

Let $\boldsymbol{P(k)}$ stand for the probability that a randomly chosen vertex has $\boldsymbol{k}$ edges attached to it. A complex network is said to follow a power law if $\boldsymbol{P(k) \propto k^{-\gamma}}$.

The parameter $\boldsymbol{\gamma}$ has a convenient approximation given in (Clauset et al. 2009):

$$\boldsymbol{\gamma \cong 1 + n\left(\sum_{i=1}^{n} ln\left(\frac{d_i}{k_{min} - 0.5}\right)\right)^{-1}} \tag{8}$$

where $d_i$ is the degree of the $i^{th}$ vertex and $k_{min}$ is the minimum degree. The Kolmogorov-Smirnov (KS) test (Clauset et al. 2009) was employed to distinguish between the following hypothesis

$H_0$: The degree distribution follows the power law distribution.

$H_1$: The degree distribution does not follow the power law distribution.

A high $p$-value suggests a good power-law fit (with exponent $\gamma$) and a low $p$-value is indicative of a poor fit.

The validity of the power law in the underlying networks has also been tested using a regression based model proposed in (Contreras-Reyes 2021) and explained as follows;

We can re-write the power-law equation as:

$$\log(P(k)) \propto -\gamma \log(k)$$

The slope $-\gamma$ is computed via the Ordinary least squares method.

Let $H_0$ stand for the null hypothesis that the data follows the power law distribution, and $H_1$ is the alternate hypothesis that the data does not follow the power law distribution.

The regression based model distinguishes between $H_0$ and $H_1$ on the basis of $R^2$ (0.942) – wherein, a high $R^2$ value suggests a good power-law fit (with exponent $\gamma$) and a low $R^2$ is indicative of a poor fit. Thus, this finding is consistent with the finding of the above discussed KS test.

## 3.2 Fast Newman community detection algorithm

The classical fast Newman community algorithm (Newman 2004) iteratively produces the partitions of the network vertices into communities and assesses the modularity $Q$ at each iteration. We compute $Q$ as follows:

$$Q = \sum_i (L_{ii} - a_i^2). \tag{9}$$

Here $L_{ij}$ is the fraction of all edges that connect vertices of group $i$ to those of group $j$, and $a_i = \sum_j L_{ij}$. The numbers $L_{ij}$ and $a_i$ depend on the edge weights. For, further details regarding equation 9 refer to (Newman 2004)

The algorithm begins by treating the $N$ vertices as one member communities. Next, consider all possible pairs of communities and compute the change in modularity given by $\Delta Q = L_{ij} + L_{ji} - 2a_i a_j$ for each pair. Now, merge the pair yielding the largest $\Delta Q$. Continue this process till there is no significant increase in $\Delta Q$.
In this paper we have applied the fast Newman algorithm to embedded networks in the Poincaré disc by choosing the edge weight between vertices $i$ and $j$ as follows:

$$w_{ij} = w_{ji} = \frac{1}{1+x_{ij}}. \tag{10}$$

Here $x_{ij}$ is the hyperbolic distance between the vertices $i$ and $j$ as defined in section 3. For, further details regarding equation 10 refer to (Muscoloni et al. 2017; Kovács and Palla 2021).

# 3.3 Coalescent embedding algorithm

Coalescent embedding is recent technique introduced in (Muscoloni et al. 2017) and has non-linear dimensionality reduction at its core. The dimensionality reduction is usually accomplished through popular algorithms like Isomap (ISO) (Tenenbaum et al. 2000), Non-centred Isomap (ncISO), Laplacian eigenmaps (LE) (Belkin and Niyogi 2003), Minimum curvilinear embedding (MCE) and Non-centred minimum curvilinear embedding (ncMCE) (Cannistraci et al. 2010; Cannistraci et al. 2013). All these algorithms embed the weighted network (with weighted adjacency matrix $W$) into the $d$-dimensional Euclidean space. In this paper, we choose the elements of $W$ as $w_{ij} = \rho_{ij}$ if vertex $i$ and $j$ are adjacent; and 0 otherwise, and choose $d = 2$. The matrix $W$ is of order $N \times N$ and thus at this point, we have, for each vertex $i$, a pair of numbers $(x_i, y_i)$. Thus we can assign an angular coordinate $\theta_i = \tan^{-1}\left(\frac{y_i}{x_i}\right)$ corresponding to each vertex. These angular values are referred to as circular adjustment (CA). The equidistance adjustment (EA) for the $i^{th}$ vertex is defined by

$$\theta_i{}' = \frac{2\pi}{N}(t_i - 1).$$

Here $t_i$ is the rank assigned to the $i^{th}$ vertex obtained after sorting the original angular coordinates in ascending order.

In order to assign the radial coordinate to each vertex in the network, the vertex degrees are sorted in descending order $d_1 \leq d_2 \leq \cdots \leq d_N$ (i.e., highest degree vertices appearing first). Then radial coordinate for each $i = 1, 2, \ldots, N$ is chosen as:

$$r_i = \frac{2}{\zeta}[\beta \ln(i) + (1-\beta)\ln(N)]. \tag{11}$$

Thus, each vertex of the network can be represented by $(r_i, \theta_i)$ and the above expression of $r_i$ forces each vertex to belong to the hyperbolic space. Also, in view of (Dorogovtsev et al. 2000), the above transformation forces the embedded vertices to follow a power law with exponent $\beta = \frac{1}{\gamma-1} \in (0,1]$ as $2 < \gamma < 3$. For, further details regarding the equation 11 refer to (Muscoloni et al. 2017; Papadopoulos et al. 2012).

# 3.4 Embeddings into Euclidean Space of lower dimension

Assume that an observed data set $X = \{x_1, x_2, \ldots, x_N\} \subseteq \mathbb{R}^D$ be a set of $N$ data points lie on an unknown manifold $M \subseteq \mathbb{R}^D$. The goal is to project the data points onto a Euclidean space of dimension $d \ll D$ in a meaningful way. A general strategy to do this is suggested in (Wang and Wang 2012) wherein a coordinate mapping $f: M \to \mathbb{R}^d$ preserving the geodesics distance is found. The set $Y = \{f(x_1), \dots, f(x_N)\}$ is the required low dimensional representation of $X$. In practice, the Isomap algorithm is a popular method for accomplishing this task ( Nie and Song 2018; Tenenbaum et al. 2000; Liu et al. 2012).

## 3.5 Statistical Measures

In this section we briefly discuss various well-known statistical methods used to capture the similarities between the clusters and the communities. The first two measures are Normalised Mutual Information (NMI) and Adjusted mutual Information (AMI) introduced in (Danon et al. 2005; Danon et al. 2006; Vinh et al. 2009) and the third is the Angular separation index (ASI) introduced in (Muscoloni and Cannistraci 2019). The ASI is particularly useful to measure the separation between communities in the setting of Poincaré disc since its formulation depends on the angular coordinates assigned to the vertices. To be precise the ASI score is computed between two "most distant" nodes in a given community. Let the observed ASI be $\alpha$. We repeat the calculation of the ASI for a large number of random shuffling of the angular coordinates of the vertices in the Poincaré disc. Each shuffling yields a different ASI. Thus, calling $ASI^k$ to be the computed ASI for the $k^{th}$ shuffling we can empirically calculate the $p$-value to determine the statistical significance of the observed ASI. Mathematically this $p$-value is given by:

$$p = \frac{1+\sum_k \delta(ASI^k \geq \alpha)}{1+Z}. \tag{12}$$

where $Z$ is the total number of shuffling and $\delta(x)$ =1, $if\ ASI^k \geq \alpha$ and $\delta(x)$ =0, if $ASI^k < \alpha$. For the further details, regarding the equation 12 refer to (Muscoloni and Cannistraci 2019).

On the other hand, the NMI between two partitions $A = \{A_1, A_2, \dots, A_{N_1}\}$ and $B = \{B_1, B_2, \dots, B_{N_2}\}$ of the network is defined based on mutual information (Nie and Song 2018; Contreras-Reyes 2022).

$$MI(A,B) = -\sum_{i=1}^{N_1}\sum_{i=1}^{N_2} M(i,j)\log\left(\frac{M(i,j)N}{r(M_i).\,c(M_j)}\right) \tag{13}$$

$$NMI(A,B) = \frac{-2 * \sum_{i=1}^{N_1}\sum_{i=1}^{N_2} M(i,j)\log\left(\frac{M(i,j)N}{r(M_i).\,c(M_j)}\right)}{\sum_{i=1}^{N_1} r(M_i)\log\left(\frac{r(M_i)}{N}\right) + \sum_{j=1}^{N_2} c(M_j)\log\left(\frac{c(M_j)}{N}\right)}. \tag{14}$$

Here $M(i,j)$ stands for the number of vertices common to $A_i$ and $B_j$, and $r(M_i) = \sum_j M(i,j)$, $c(M_j) = \sum_i M(i,j)$ are row sum and column sum of the confusion matrix $M$, respectively. NMI is the normalized mutual information with its values lying in the interval [0,1]. If the two partitions are exactly same, then the $NMI(A,B)$ will attain its maximum value which is 1. On the other hand, $MI(A,B) = 0$ and $NMI(A,B) = 0$ when the two network partitions are independent of each other (Contreras-Reyes 2022). Also, a high value of $NMI(A,B)$ (close to 1) will indicate that there is high degree of dependency between the network partitions.
Using the expected mutual information defined in (Vinh et al. 2009), the adjusted mutual information for the shuffled network partitions $A'$ and $B'$, which addresses the issue of random chances and can be expressed as follows:

$$AMI(A,B) = \frac{MI(A,B) - E(MI(A',B'))}{\frac{1}{2}\left(\sum_{i=1}^{N_1} r(M_i)\log\left(\frac{r(M_i)}{N}\right) + \sum_{j=1}^{N_2} c(M_j)\log\left(\frac{c(M_j)}{N}\right)\right) - E(MI(A',B'))} \tag{15}$$

For, further details regarding the equation 15 refer to (Vinh et al. 2009).

# 4 Methodology

Our findings establish the superiority of embedding the networks in a hyperbolic space (coalescent embedding) versus embedding same networks in the Euclidean space on the following four counts:

- Topological communities versus clusters of the embedded networks.
- Identifying the periods of the volatility and stability of the market.
- Modularity of embedded network – a tool to capture market sensitivity.
- Visualizing Market sectors through the coalescent embedding.

We elaborate on each of the above questions in sections 4.1-4.4 given below.

## 4.1 Topological communities versus clusters of the embedded networks

**Step 1.** Define a pairwise correlation-based network of stocks using the available time series data in the section 2.

We define a correlation based network represented by $G = (V, E)$ from the data where the set of vertices $V$ consists of the 383 stocks and the weight assigned to each edge $e_{ij}$ is the Pearson correlation coefficient $\rho_{ij}$ given by

$$\rho_{ij} = \frac{E[r_i r_j] - E[r_i]E[r_j]}{\sqrt{E[r_i^2] - E[r_i]^2}\sqrt{E\left[r_j^2\right] - E[r_j]^2}}. \quad (16)$$

Here $r_i$ and $r_j$ is the corresponding log returns of stocks $i$ and $j$, respectively. This construction is in the spirit of (Jiang et al. 2014; Millington and Niranjan 2017; Sharma and Habib 2019; Guo et al. 2018; Aste et al. 2010; Pozzi et al. 2013; Chi et al. 2010; Kumar and Deo 2012; Tumminello et al. 2005; Nie and Song 2018).

**Step 2.** Extract subgraphs by eliminating redundant edges to mitigate the effect of noise (Jiang et al. 2014; Millington and Niranjan 2017; Sharma and Habib 2019; Guo et al. 2018; Aste et al. 2010; Pozzi et al. 2013; Chi et al. 2010; Kumar and Deo 2012; Tumminello et al. 2005; Nie and Song 2018) and introduce heterogeneity in the complex system which will justify the embedding into the hyperbolic space according to preliminary section 3. The sub graphs that we worked with were the MST and the PMFG. The MST is extracted using the Prim's algorithm (Prim 1957) and the PMFG was extracted using the procedure outlined in preliminary section.

- We carried out the community analysis using the fast Newman algorithm (Newman 2004).
- The communities extracted in the previous step will be referred to as topological communities and the collection of all topological communities is denoted by $C^{top}$.

**Step 3.** Using the procedure outlined in section 3.1, the validity of the power law was examined for the MST, PMFG and PTN subgraphs. In case of MST (obtained from the full correlation network) we choose $k_{min} = 2$ and choose $k_{min} = 3$ in case of PMFG. High $p$-values ($> 0.1$) in both cases indicate the presence of power law distribution. However no conclusive evidence of power law distribution was seen in case of PTN subgraphs and hence we discard PTN altogether from our further analysis. The respective estimates of $\gamma$ and the $p$-values for the MST, PMFG and PTN networks corresponding yearly data is provided in the supporting information (S1 Table). It has also been observed (by repeating the hypothesis test on yearly data) that the networks constructed from 60 day windows also follow the Power law distribution. For the sake of brevity we present the results for MST only in this paper and relegate those of PMFG in the supporting information since the conclusions are very similar.

**Step 4.** Carry out the coalescent embedding for the corresponding sub graphs (obtained in step 2) to embed them in Poincaré disc. This identifies each vertex $i$ of the graph with a pair of numbers $(r_i, \theta_i)$ as described in section 3.3.

- Compute the clusters in the Poincaré disc using the hyperbolic version of k-means and referred as $C^{hyp}$.

**Step 5.** Similarly to step 4, carry out the Euclidean embeddings for the corresponding sub graphs (obtained in step 2). This identifies each vertex $i$ of the graph with a pair of numbers $(x_i, y_i)$ in Euclidean plane as described in section 3.4.

- Compute the clusters in the Euclidean space using k-means and referred as $C^{Euc}$.

**Step 6.** Compare the quality of topological communities with the low dimensional Euclidean clusters and hyperbolic clusters on the basis of NMI (or AMI) score.

## 4.2 Identifying the periods of the volatility and stability of the market

**Step 1.** To evaluate the geometrical changes in the stock networks subject to two different Market conditions (Healthy period and crisis period), we selected two specific years data among the five years data. For the healthy period we selected year 2018 characterized by low volatility and for the crisis period year 2020 characterized by high volatility that could be seen in The Table 1 and Fig.1.

**Table 1** Price statistics of CNX100 Index of the year 2020 and 2018

| Years | Min | Mean | Max | Standard deviation |
|---|---|---|---|---|
| 2018 | 10307.70 | 11045.40 | 12028.29 | 367.52 |
| 2020 | 7719.10 | 11326.20 | 14090.75 | 1388.96 |

The table gives the minimum, maximum, mean and the standard deviation of the prices of CNX100 corresponding to the year 2020 and 2018.

We also list some of the network topological properties of both periods below in Table 2.

**Table 2** Healthy period and Crisis period MST networks topological properties

| Years | Weighted mean degree (MD) | Average shortest path distance (SD) | Average local clustering coefficient (CC) | Average edge weight (EW) |
|---|---|---|---|---|
| 2018 | 0.9292 | 6.0995 | 0 | 0.4658 |
| 2020 | 1.0210 | 7.7720 | 0 | 0.5118 |

This table gives the Mean of weighted degree, Average shortest path distance and Average edge weight Healthy period (2018) and the Crisis period (2020).

**Step 2.** Following the step 1 to 4 of section 4.1, we, generate the series of correlation-based networks (MST, PMFG) with the a moving window of size 60 days and the sliding window for 1 day over both periods. The total number of windows were 137 and 124 in Year 2020 and Year 2018, respectively. Next, we embedded the generated networks in the Poincaré disc via various class of the coalescent embedding algorithm. Further, for each network, we compute the numbers HD (equation 3), HSD for the embedded networks and average edge weight (EW), average shortest path distance (SD) for the original networks. Computation of HSD involves finding the shortest path between pairs of vertices (usually by Dijkstra's algorithm (Dijkstra 2022)) with edge weights chosen to be the hyperbolic distance between the vertices and then summing up the edge weights.

**Step 3.** The periods of volatility are distinguished from the periods of stability on the basis of the Mann Whitney test (Zar 1999) applied on either the HD values or the HSD values of the networks in the two categories.

## 4.3 Modularity of embedded network – a tool to capture market sensitivity.

**Step 1.** Similar to step 2 refer in section 4.2, we generate the series of correlation-based MST networks with the calculation window for 60 days and the sliding window for 1 day over the five years dataset. There were total 879 windows. We also embedded each of these networks in Poincaré disc via various class of coalescent embedding algorithm.

**Step 2.** Next, we use the Newman fast community detection algorithm over each of the above 879 original networks and embedded networks, respectively. Then we compute the communities and corresponding modularity value (equation 9). We refer these modularity's as original modularity and the hyperbolic modularity, respectively. For the embedded networks, we use edge weights as defined in (equation 10).

**Step 3.** Now setup two time series $\{O_t\}_{t=1}^{T}$ and $\{H_t\}_{t=1}^{T}$, $T = 1,2,3,\ldots,859$ as original modularity of the original network and the hyperbolic modularity of the embedded network, respectively.

**Step 4.** We compute the Simple moving average (SMA) time series for $O_t$ and $H_t$ for $N = 20$ days as follows: $O_t' = \frac{1}{20}\sum_{j=0}^{19} O_{t-j}$ and $H_t' = \frac{1}{20}\sum_{j=0}^{19} H_{t-j}$ . The Bollinger Bands (BB) (Bollinger 2002) of $O_t'$ and $H_t'$ were calculated as $[O_t' - 3\sigma_t^o, O_t' + 3\sigma_t^o]$, and $[H_t' - 3\sigma_t^H,\ H_t' + 3\sigma_t^H]$, respectively. In the corresponding intervals the series $O_t' - 3\sigma_t^o$ and $H_t' - 3\sigma_t^H$ are called lower Bollinger Band referred as lowerBB, similarly the series $O_t' + 3\sigma_t^o$ and $H_t' + 3\sigma_t^H$ are called upper Bollinger Band referred as upperBB. Here $\sigma_t^o$ and $\sigma_t^H$ stand for the standard deviations time series for $O_t$ and $H_t$. For the sake of comparison, we also construct the BB bands for the time series of prices $\{P_t\}_{t=1}^{T}$ over the same five years period. We present the inferences drawn on the basis of the above analysis in section 5.3.

## 4.4 Visualising Market sectors through the coalescent embedding

**Step 1.** Following the steps 1 to 4 of section 4.1, we visualise certain market sectors via coalescent embedding in Poincaré disc. For this we selected stocks over five-years 2017 to 2021 from the sectors: (i) Finance (ii) Healthcare (iii) Information Technology.
**Step 2.** We create the visualisation of MST networks in the Poincaré disc corresponding to finance, healthcare sectors and all three sectors, respectively.
**Step 3.** A quantitively separation of topological communities ($C^{top}$) performed on the basis of ASI score (section 3.5) has been performed in the Poincaré disc.

# 5 Results

## 5.1 Topological communities versus clusters of the embedded networks

This analysis is performed on the data corresponding to each year 2017-2021 separately.
For each network, we carried out the steps explained in section 4.1 to generate the $C^{top}$, $C^{Euc}$ and $C^{hyp}$. Table 3 and Table 4 reports the NMI and AMI score, respectively between the $C^{top}$, $C^{Euc}$ and $C^{top}$, $C^{hyp}$. Clearly, LE-EA class of the coalescent embedding has the best NMI (or AMI) score consistently over the years in comparison to other classes. Note that the Minimum score of $NMI(C^{top}, C^{hyp})$ is 0.579 while the maximum score of $NMI(C^{top}, C^{hyp})$ is 0.854. From Table 3 and Table 4 we observe that the coalescent embedding is able to match the topological communities with hyperbolic clusters more closely than Euclidean clusters. Similar, results are observed for the PMFG networks provided in the supporting information (S2 Table and S3 Table). An example of embedding the topological communities in the Euclidean plane and the hyperbolic disc is given in Fig.2 for the MST network corresponding to year 2017 data. Different colours in each subfigure represent different communities in $C^{top}$.

**Table 3** NMI Scores: $NMI(C^{top}, C^{Euc})$ and $NMI(C^{top}, C^{hyp})$

| Methods | 2017 | 2018 | 2019 | 2020 | 2021 |
|---|---|---|---|---|---|
| ISO | 0.693 | 0.695 | 0.683 | 0.746 | 0.779 |
| ncISO | 0.608 | 0.583 | 0.649 | 0.659 | 0.688 |
| LE | 0.141 | 0.116 | 0.135 | 0.121 | 0.148 |
| MCE | 0.694 | 0.694 | 0.676 | 0.751 | 0.78 |
| ncMCE | 0.604 | 0.584 | 0.644 | 0.66 | 0.693 |
| Coal-ISO-EA | 0.705 | 0.677 | 0.641 | 0.683 | 0.826 |
| Coal-ncISO-EA | 0.698 | 0.579 | 0.644 | 0.668 | 0.797 |
| Coal-LE-EA | **0.795** | **0.772** | **0.752** | **0.817** | **0.854** |
| Coal-MCE-EA | 0.592 | 0.602 | 0.631 | 0.639 | 0.685 |
| Coal-ncMCE-EA | 0.619 | 0.72 | 0.702 | 0.791 | 0.811 |

The rows 2 to 6 of the Table 3 gives the NMI score between the $C^{top}$ and $C^{Euc}$, and next rows 7 to 11 gives the NMI score between the $C^{top}$ and $C^{hyp}$, corresponding to the MST networks respectively. EA is equidistance adjustment.

**Table 4** AMI Scores: $AMI(\mathcal{C}^{top}, \mathcal{C}^{Euc})$ and $AMI(\mathcal{C}^{top}, \mathcal{C}^{hyp})$

| Methods | 2017 | 2018 | 2019 | 2020 | 2021 |
|---|---|---|---|---|---|
| ISO | 0.625 | 0.653 | 0.623 | 0.702 | 0.744 |
| ncISO | 0.519 | 0.525 | 0.582 | 0.598 | 0.638 |
| LE | 0.005 | 0.017 | 0.019 | 0.023 | 0.041 |
| MCE | 0.628 | 0.654 | 0.615 | 0.707 | 0.746 |
| ncMCE | 0.516 | 0.528 | 0.575 | 0.6 | 0.643 |
| Coal-ISO-EA | 0.637 | 0.632 | 0.57 | 0.625 | 0.799 |
| Coal-ncISO-EA | 0.628 | 0.522 | 0.574 | 0.607 | 0.764 |
| Coal-LE-EA | **0.748** | **0.741** | **0.703** | **0.784** | **0.831** |
| Coal-MCE-EA | 0.502 | 0.547 | 0.558 | 0.575 | 0.634 |
| Coal-ncMCE-EA | 0.535 | 0.681 | 0.643 | 0.752 | 0.78 |

The rows 2 to 6 of the Table 4 gives the AMI score between the $\mathcal{C}^{top}$ and $\mathcal{C}^{Euc}$, and next rows 7 to 11 gives the AMI score between the $\mathcal{C}^{top}$ and $\mathcal{C}^{hyp}$, corresponding to the MST networks respectively. EA is equidistance adjustment.

Further, we statistically show that the NMI's of $\mathcal{C}^{hyp}$ with $\mathcal{C}^{top}$ ($NMI_{top}^{hyp}$) is consistently greater than NMI of $\mathcal{C}^{Euc}$ with $\mathcal{C}^{top}$ ($NMI_{top}^{Euc}$). We calculate the NMI's for each of the networks generated by rolling windows of size 60 days rolled over by a day. There were total 879 window's over the five years 2017-2021. Mann Whitney test is a non-parametric test, which test the variabilities between two sub populations. We used this test to check that $NMI_{top}^{hyp}$ values are more than $NMI_{top}^{Euc}$ values. The $H_0$ stand for the null hypothesis that the $NMI_{top}^{hyp}$ values are not greater than those $NMI_{top}^{Euc}$ values, and $H_1$ is the alternate hypothesis that the $NMI_{top}^{hyp}$ values are greater than those $NMI_{top}^{Euc}$ values. With reference to the $p$-values listed in the table 5, we reject the $H_0$ and accept the $H_1$ with 0.05 significance level corresponding to the three classes of the coalescent embedding algorithm. Clearly, on the basis of statistical analysis we can conclude that the coalescent embedding methods matches topological communities more closely with hyperbolic clusters than Euclidean clusters. Similar results are observed in the PMFG network provided in supporting information (S4 Table).

**Table 5** NMI dynamics of $NMI_{top}^{hyp}$ and $NMI_{top}^{Euc}$ for MST networks

| Methods | mean $NMI_{top}^{hyp}$ | mean $NMI_{top}^{Euc}$ | $p$-value |
|---|---|---|---|
| ISO, Coal-ISO-EA | 0.654 | 0.698 | 1 |
| ncISO, Coal-ncISO-EA | **0.655** | **0.594** | **3.27297878e-75** |
| LE, Coal-LE-EA | **0.766** | **0.121** | **7.86206154e-289** |
| MCE, Coal-MCE-EA | 0.585 | 0.698 | 1 |
| ncMCE, Coal-ncMCE-EA | **0.659** | **0.595** | **8.36831297e-77** |

The table 5, gives the mean of the NMI calculated over the 879 windows, comparing $\mathcal{C}^{hyp}$ with $\mathcal{C}^{top}$ ($NMI_{top}^{hyp}$) and NMI of $\mathcal{C}^{Euc}$ with $\mathcal{C}^{top}$ ($NMI_{top}^{Euc}$) for each class of the coalescent embedding algorithm. Additionally, the table also gives the $p$-values obtained from the MW test, which are significant at the level of 0.05. Note, that the mean of $NMI_{top}^{hyp}$ is greater than the mean of $NMI_{top}^{Euc}$ across the three classes of the coalescent embedding.

## 5.2 Identifying the periods of the volatility and stability of the market

As explained in the section 4.2 step 2, moving window networks corresponding to the years 2018 and 2020 were embedded into Poincaré disc using the coalescent embedding algorithm. Table 6 reports, for each method (EA and CA), mean of the geometrical measures HD and HSD for the two years along with the $p$-values of the Mann-Whitney test, where $H_0$ stand for the null hypothesis that the mean measure of networks corresponding to year 2020 is not greater than the mean measure of networks corresponding to year 2018, and the $H_1$ stand for the alternative hypothesis that mean measure of networks corresponding to year 2020 is greater than the mean measure of networks corresponding to year 2018. For reference, we also compute the mean of SD, and EW measures. We wish to point out that considering the mean of these measures is appropriate in light of the analysis presented below which shows that the distributions of SD, EW, HD, and HSD do not follow a power law distribution at the 0.01 level of significance. In Fig.3 and Fig.4 the probability distributions of measures plotted as histogram corresponding to MST networks of both the periods, respectively and for PMFG provided in the supporting information (S1 Fig and S2 Fig). Note that in Table 6, for the case of SD measure of original networks, $p$-values $< 0.05$, leads to reject the null hypothesis $H_0$ and accept the alternative $H_1$ at 0.05 level of significance, but not in the case of EW measure. On the other hand, the $p$-values in case of HD and HSD for all the classes of coalescent embedding methods are less than 0.05, and even equal to zero in some cases rejecting the null and accepting the alternative at 0.05 level of significance, which indicates that the Mann Witney test on SD measure is also able to segregate the both periods but hyperbolic measures HD and HSD are able to segregate more effectively. Similar, results are observed for the PMFG networks provided in the supporting information (S5 Table).

**Table 6** Healthy period versus Crisis period: Mann Whitney Test $p$-values for MST networks

| Methods | mean measure crisis period (year 2020) | mean measure healthy period (year 2018) | $p$-value |
|---|---|---|---|
| ISO-CA-HD | 20.13 | 19.77 | 2.71E-06 |
| ISO-CA-HSD | 152.80 | 115.41 | 0.00E+00 |
| ISO-EA-HD | 21.311 | 21.03 | 8.40E-10 |
| ISO-EA-HSD | 180.39 | 140.20 | 0.00E+00 |
| ncISO- CA-HD | 20.41 | 20.03 | 1.18E-07 |
| ncISO- CA-HSD | 164.36 | 124.92 | 0.00E+00 |
| ncISO-EA-HD | 21.31 | 21.03 | 8.44E-10 |
| ncISO-EA-HSD | 180.36 | 141.05 | 0.00E+00 |
| LE- CA-HD | 20.13 | 19.62 | 3.40E-08 |
| LE- CA-HSD | 138.29 | 99.01 | 0.00E+00 |
| LE-EA-HD | 21.311 | 21.03 | 8.57E-10 |
| LE-EA-HSD | 173.36 | 133.80 | 0.00E+00 |
| MCE- CA-HD | 20.44 | 20.15 | 4.12E-10 |
| MCE- CA-HSD | 173.57 | 138.25 | 0.00E+00 |
| MCE-EA-HD | 21.31 | 21.03 | 8.44E-10 |
| MCE-EA-HSD | 182.64 | 144.34 | 0.00E+00 |
| ncMCE- CA-HD | 20.44 | 20.03 | 6.29E-08 |
| ncMCE- CA-HSD | 173.57 | 131.94 | 0.00E+00 |
| ncMCE-EA-HD | 21.31 | 21.03 | 8.53E-10 |
| ncMCE-EA-HSD | 182.64 | 142.48 | 0.00E+00 |
| **Average EW** | **0.2879** | **0.2759** | **0.2633** |
| **Average SD** | **9.6609** | **8.0407** | **1.11E-14** |

Table 6 gives the mean values of Hyperbolic Distance (HD) and Hyperbolic Shortest Path Distance (HSD) of the embedded MST network constructed for both crisis and healthy periods, categorized by each class of coalescent embedding. Additionally, the table includes the mean values of Edge Weight (EW) and Shortest Path Distance (SD) for the original MST network during crisis and healthy periods. In this context, Equidistance Adjustment (EA) and Circular Adjustment (CA) are used. HD and HSD measure distances within the hyperbolic space, whereas EW and SD refer to the original network's edge weight and shortest path distance, respectively. The last column of the table reports the $p$-values obtained from the MW test for both the periods and corresponding to each class. Note that crisis period mean measures are greater than the healthy period in all cases.

## 5.3 Modularity of embedded network – a tool to capture market sensitivity

In Fig.5, Fig.6 and Fig.7 we observe that the SMA series of the hyperbolic modularity (Corresponding to ISO-CA class of Coalescent embedding algorithm) crosses its respective lowerBB even before the SMA series of the original modularity and the prices series crosses their respective lowerBB. For example the first time SMA series of hyperbolic modularity crosses its lowerBB at tick 14 (date 28/06/17) before the crossover of the SMA series of the original modularity at tick 52 (date 07/09/17) and the prices series at the tick 188 (date 23/05/18). With this we observe that the hyperbolic modularity is more sensitive to the market events, crisis or fluctuation in compared to original network modularity and captures the market trend even before captured by the original modularity. As a result, financial investor can monitor the hyperbolic modularity as an early indicator of the market movement and potential trading opportunities in future.

We also performed the one tailed Mann Whitney test in order to test the hypothesis that

$$H_0: \mu_H < \mu_O$$

$$H_A: \mu_H \geq \mu_O$$

Where $\mu_H$ is the mean of the hyperbolic modularity and the $\mu_O$ is the mean of the original modularity. We are able to reject the null hypothesis at 0.01 level of significance level, which statistically proves that the mean of the hyperbolic modularity is greater than the mean of the original modularity.

## 5.4 Visualising Market sectors through the coalescent embedding

Co-movement of stocks have been studied in the literature using techniques like Multidimensional scaling dimensionality reduction, Dendrogram and MST methods (Sharma et al. 2017), and Random Matrix Theory (Sharma and Habib 2019). Local network indices have also been a popular tool to analyse co-movement of stocks. For this, we have selected stocks over the five-years 2017 to 2021 from three sectors: (i) Finance (ii) Healthcare (iii) Information Technology. We embedded the MST network corresponding to the stocks in Finance and Healthcare sector in Fig.8 and also corresponding stocks on all three sectors Finance, Healthcare and Information technology in Fig.9. It is quite evident that coalescent embedding visually is able to segregates sectors clearly.

The separation of topological communities ($C^{top}$) quantitatively on the basis of ASI has been performed in the hyperbolic disc. Table 7 gives the ASI scores showing the degree of separation of topological communities of MST network in the 2D hyperbolic space in the context of various embedding algorithms and years. A score close to 1 ($p$-value $< 0.001$) indicate the successful separation of topological communities. The result clearly indicates that all the classes of the coalescent embedding methods are able to segregate the topological communities in the hyperbolic space with LE-EA showing the best results. For all the methods, we note that the observe $p$-values are close to $0.00099$. Similar, results are observed for the PMFG networks in the supporting information (S6 Table).

**Table 7** ASI index score for the topological communities ($C^{top}$) for MST networks

| Methods | 2017 | 2018 | 2019 | 2020 | 2021 |
|---|---|---|---|---|---|
| ISO-EA | 0.8588 | 0.8725 | 0.8449 | 0.8776 | 0.9620 |
| ncISO-EA | 0.8882 | 0.7991 | 0.8637 | 0.8720 | 0.9447 |
| LE-EA | **0.9683** | **0.9694** | **0.9490** | **0.9776** | **0.9770** |
| MCE-EA | 0.8001 | 0.8307 | 0.8520 | 0.8614 | 0.8903 |
| ncMCE-EA | 0.7750 | 0.9237 | 0.9179 | 0.9547 | 0.9671 |

The table 7, gives the yearly ASI score for the topological communities corresponding to each class of the coalescent embedding algorithm. EA represent the equidistance adjustment. Note that the LE-EA class is the clear winner.

# 6 Conclusion

In this study, we consider the Poincaré disc as a natural framework to explore the Indian stock market and demonstrate the use of the hyperbolic geometry underlying the subnetworks (MST, PMFG) of the full correlation-based network. The paper comprehensively explores the community structure of the networks via the fast version of the Newman community detection algorithm and compares it with the kmeans clusters obtained after embedding the networks in the Euclidean space as well as hyperbolic space using NMI (or AMI) scores. Results shows that some classes of the coalescent embedding like (ncISO, LE, ncMCE) outperforms the respective Euclidean embeddings methods, and LE class of coalescent embedding is clearly the winner with NMI ranging from 0.752 to 0.854. We also show statistically with $p$-value $<$ $0.05$, the hyperbolic kmeans clusters are actually more richer than the Euclidean kmeans clusters. This research also demonstrates the use of network parameters HD and HSD in order to effectively differentiate between periods of market stability and volatility. The periods are disguised based on $p$-value $< 0.05$ obtained using statistical test over the HD and HSD. Next, we showed that the hyperbolic modularity of embedded network has ability to spot significant changes in the market early. Precisely, we are able to detect significant changes in the market about 20 days in advance by applying Bollinger band analysis on the modularity of embedded networks with lower BB and upper BB fixed at 3 standard deviations away from the mean. This clear early signal of market change is not observed if we consider the modularity of the original networks. The ASI scores of the topological communities in the embedded network are seen to lie between 0.78-0.98 and this reveals that they are effectively segregated. Finally, we distinctly segregate and visualize specific market sectors within the Poincaré disc utilizing the Equidistant angular coordinate in coalescent embedding. This analysis highlights the innate clustering capability of coalescent embedding. This study is applicable in various real-world domains, provided that the underlying network degree distribution is heterogeneous scale free in nature. There are real world domains like, epidemiology, biology, engineering etc, where the underlying network is commonly heterogenous, thus this study further can be employed to understand the systems. In summary, this research amalgamates financial analysis, network science, hyperbolic geometry, and machine learning to shed new and illuminating insights on the dynamics of the Indian stock market. It introduces an innovative approach that effectively distinguishes market volatility from stability, enabling the early detection of market shifts. Furthermore, it showcases the superior performance of hyperbolic embedding in comprehending complex financial systems. In future, In order to see the consistency of this work other than Indian stock market, we will also perform this study in other foreign stock market for example new York stock, Shanghai stock exchange etc. Also, the hyperbolic clusters can be leveraged to create new portfolios and it would be interesting to analyse their properties in terms of rate of return and risk.

# Acknowledgement

We thank the Shiv Nadar Institution of Eminence for providing the computational facilities and the necessary infrastructure needed to carry out the present research. We extend a special gratitude to Professor Sanjeev Agrawal for his encouragement and valuable comments. The first author is thankful to the Council of Scientific & Industrial Research (CSIR), India, Senior Research fellowship scheme (Grant No. 09/1128(12964)/2021-EMR-I) for financial support for this research.

# Statements & Declarations

**Funding :** The authors declare that no funds, grants, or other support were received.

**Competing Interests:** The authors declare no competing financial interests.

**Author Contributions:** All authors contributed to the conceptualisation, formal analysis, simulation and Manuscript preparation.

**Code and Data Availability:** The code and data that support the findings of this study are publicly available at the GitHub repository: https://github.com/ML-GM/Market-geometry-

# Supporting Information

**S1 Fig Probability distribution of the mean measures of PMFG networks corresponding to healthy period**
**S2 Fig Probability distribution of the mean measures of PMFG networks corresponding to crisis period**
**S1 Table Estimated power law exponent $\gamma$ of the degree distribution of PMFG networks**
**S2 Table NMI Scores: $NMI(\mathcal{C}^{top}, \mathcal{C}^{Euc})$ and $NMI(\mathcal{C}^{top}, \mathcal{C}^{hyp})$ for PMFG networks**
**S3 Table AMI Scores: $AMI(\mathcal{C}^{top}, \mathcal{C}^{Euc})$ and $AMI(\mathcal{C}^{top}, \mathcal{C}^{hyp})$ for PMFG networks**
**S4 Table NMI dynamics of $NMI_{top}^{hyp}$ and $NMI_{top}^{Euc}$ for PMFG networks**
**S5 Table Healthy period versus Crisis period: Mann Whitney Test $p$-values for PMFG networks**
**S6 Table ASI Score for topological communities of PMFG networks**

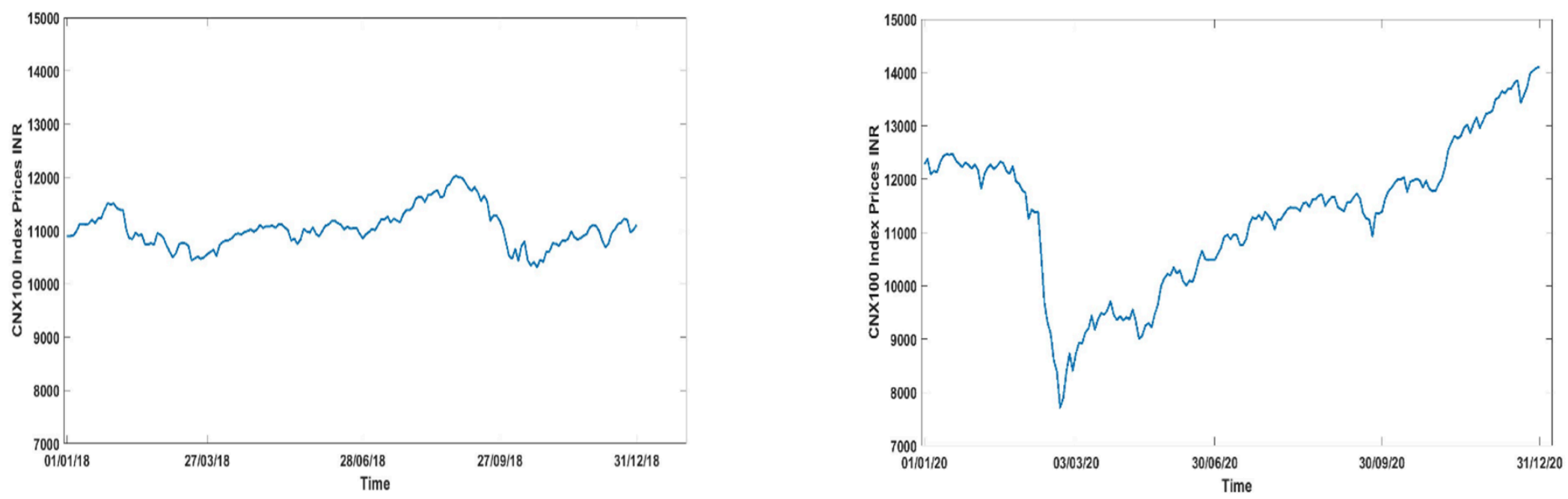

(a) Year 2018

(b) Year 2020

**Fig. 1** Price trend of CNX100 index for (a) Year 2018 (b) Year 2020. A sharp fluctuation is noted for year 2020.

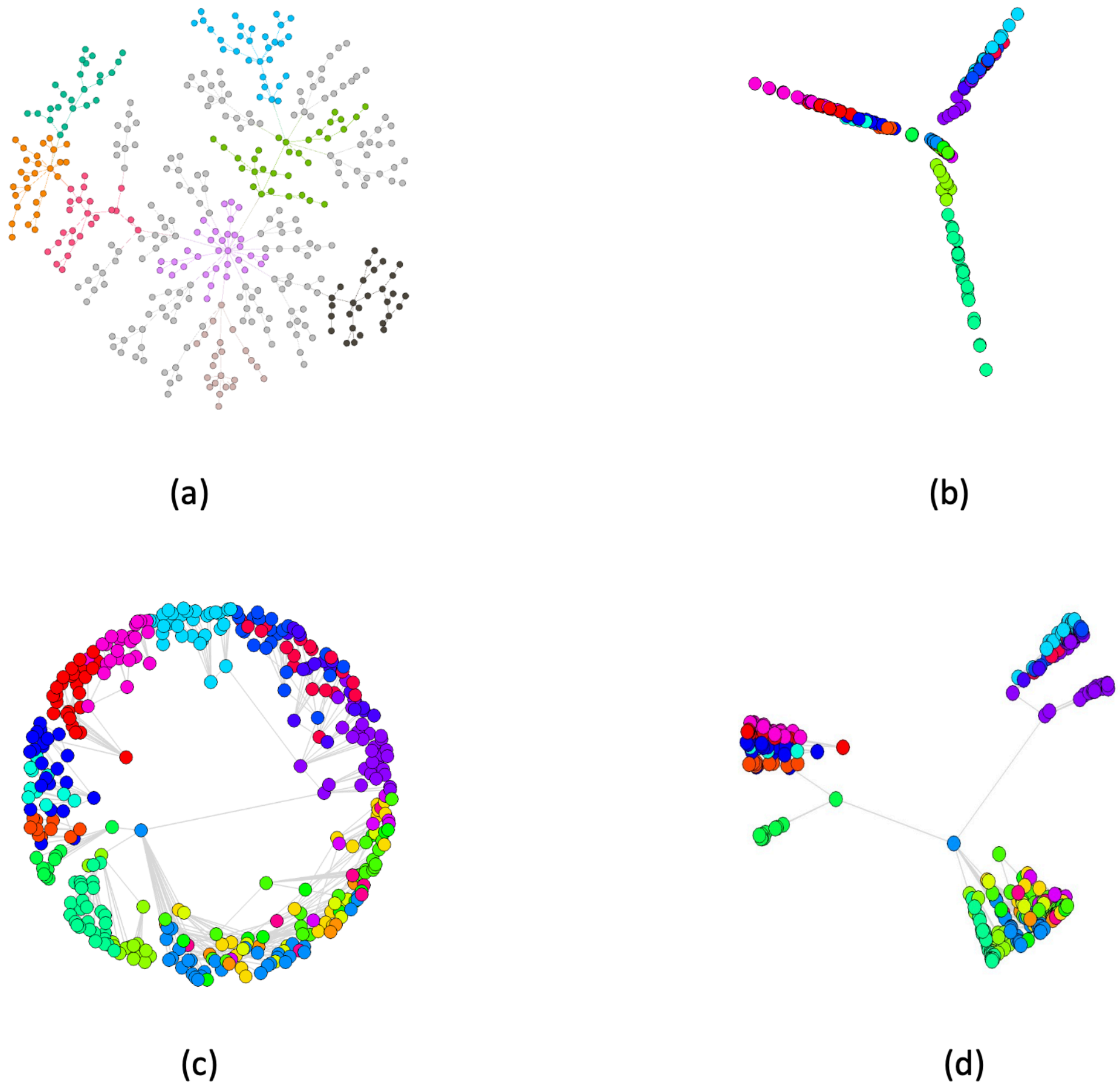

**Fig. 2** Networks constructed from 2017 year data. a) MST network b) MST network embedded in Euclidean plane via Isomap c) MST network embedded in Poincaré disc via coalescent embedding (ISO-EA) d) MST network embedded in Poincaré disc via coalescent embedding (ISO-CA). Various communities given by Newman fast algorithm are painted in different colours. Communities appears to be more distinguished in Fig.2(c).

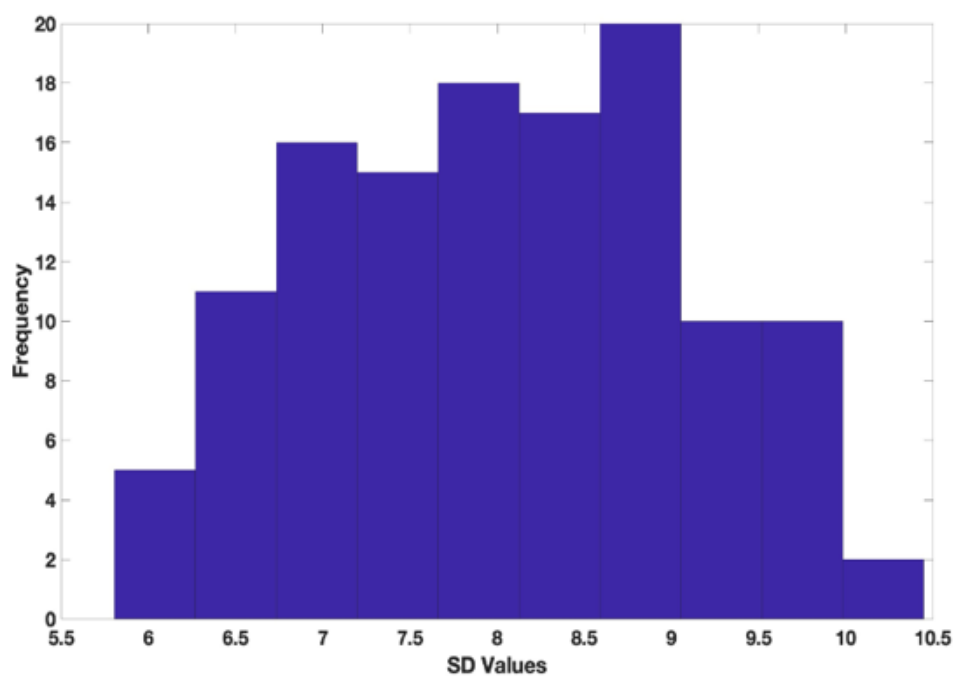

a) SD

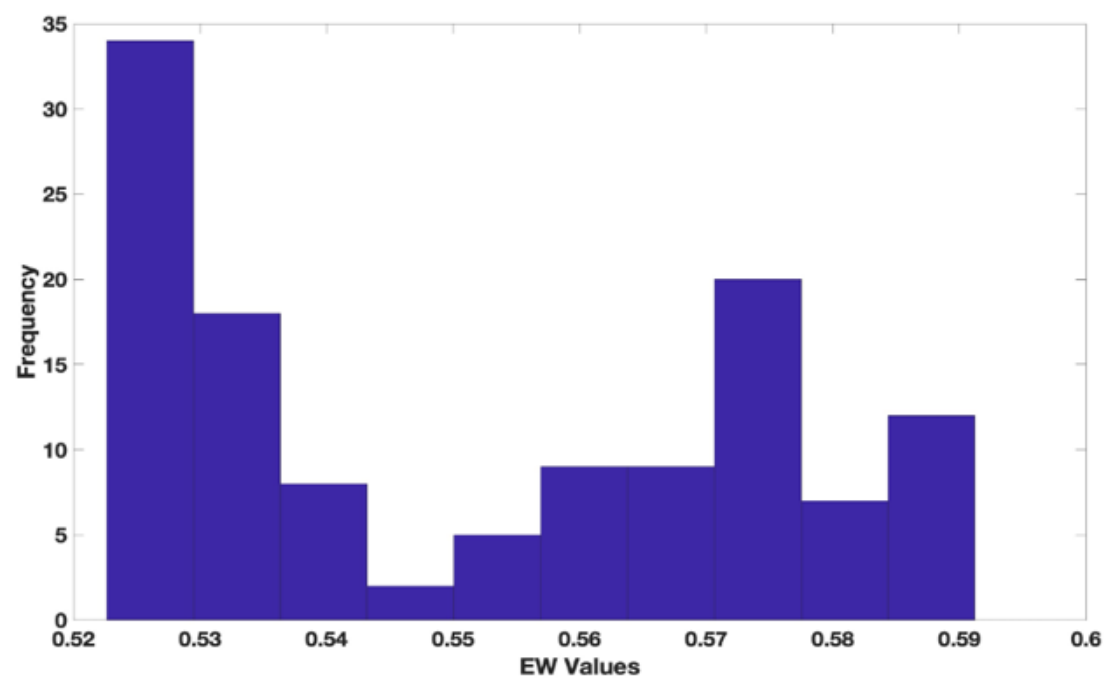

b) EW

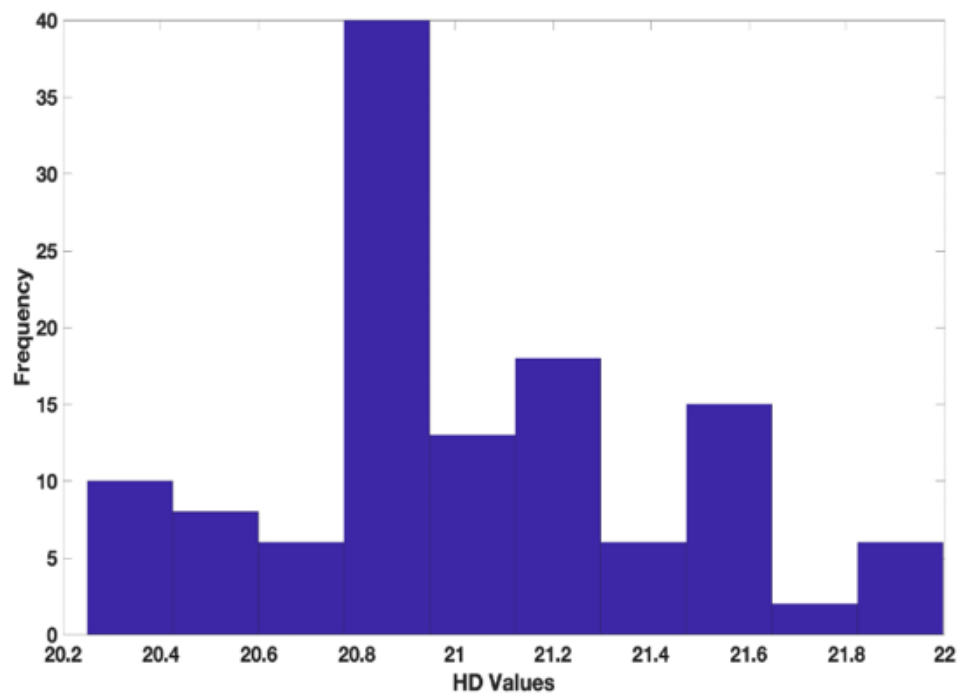

c) HD

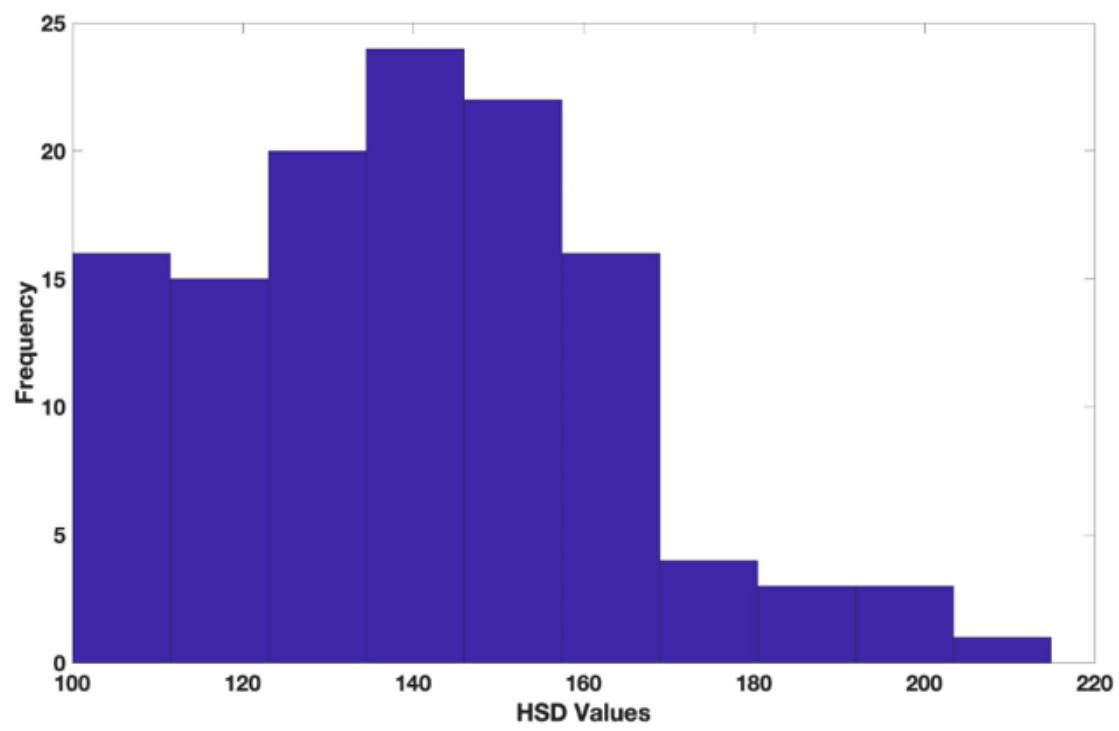

d) HSD

**Fig. 3** Probability distribution of the mean measures a) SD, b) EW of the original network and mean measures c) HD, d) HSD of embedded network (ISO-EA) for healthy period year 2018. The figures are suggestive that the various measures do not obey a power law distribution.

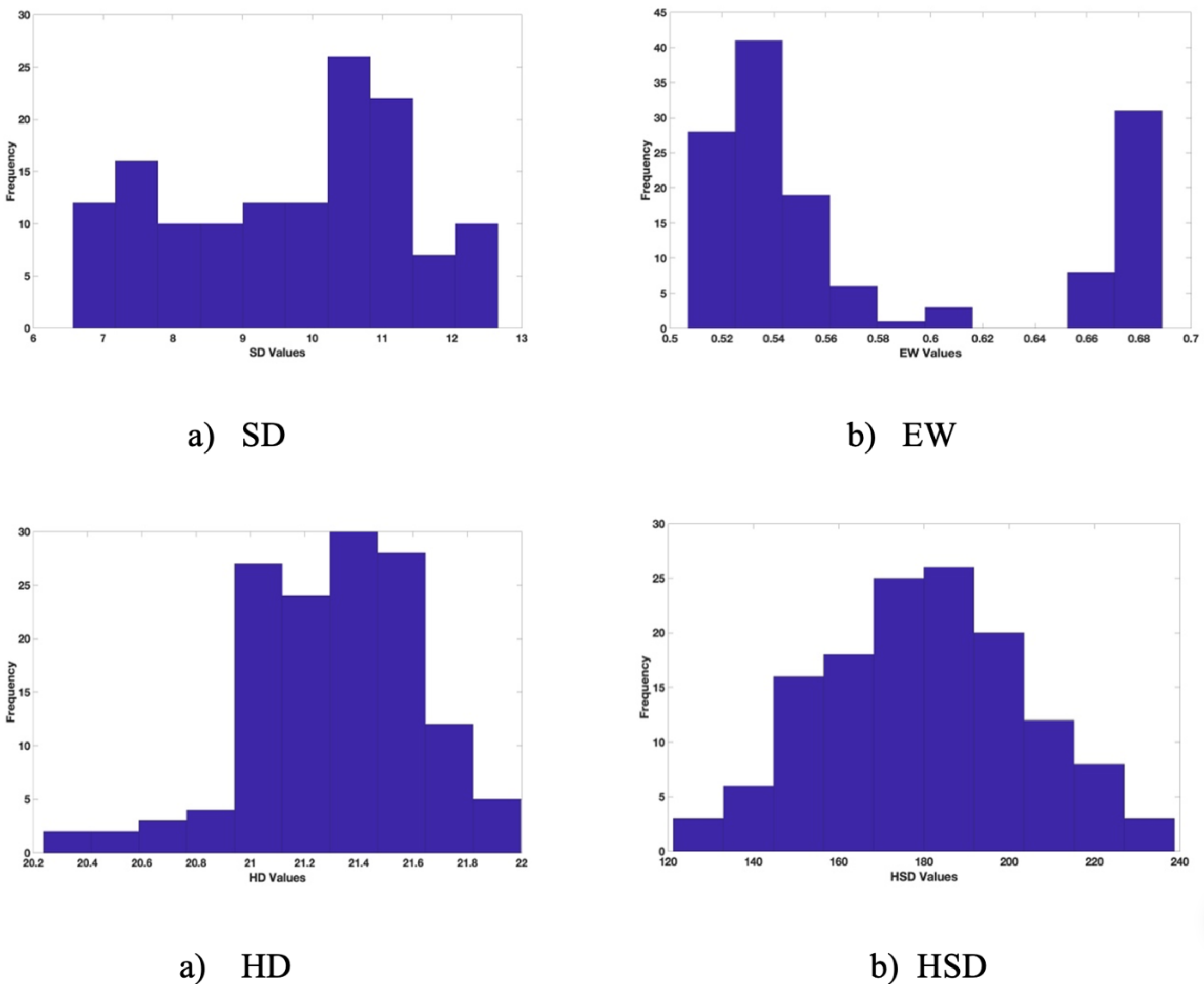

**Fig. 4** Probability distribution of the mean measures a) SD, b) EW of the original network and mean measures c) HD, d) HSD of embedded network (ISO-EA) for healthy period year 2020. The figures are suggestive that the various measures do not obey a power law distribution.

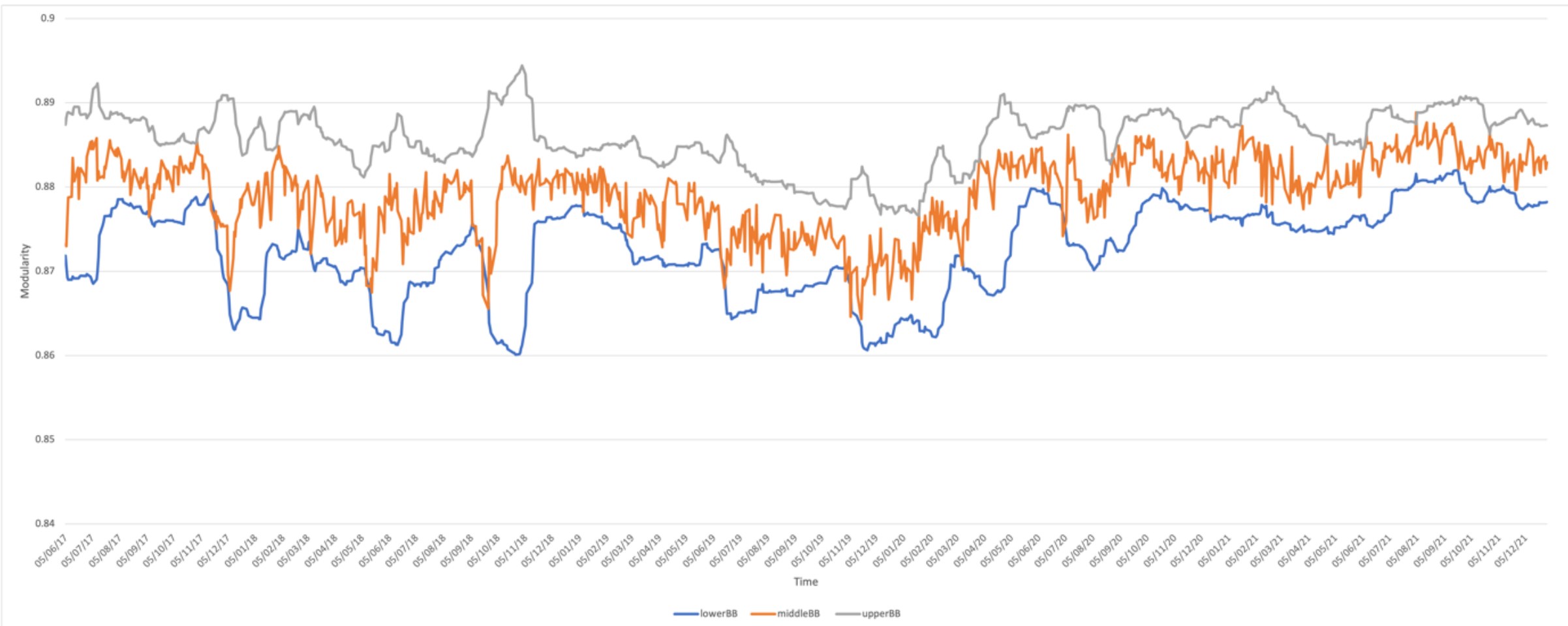

**Fig. 5** Bollinger Bands on the time series of original modularity**;** The orange curve is middleBB denotes the 20-day SMA series; gray curve and the blue curve are upperBB and lowerBB, respectively denoting 3 standard deviations above and below the SMA series.

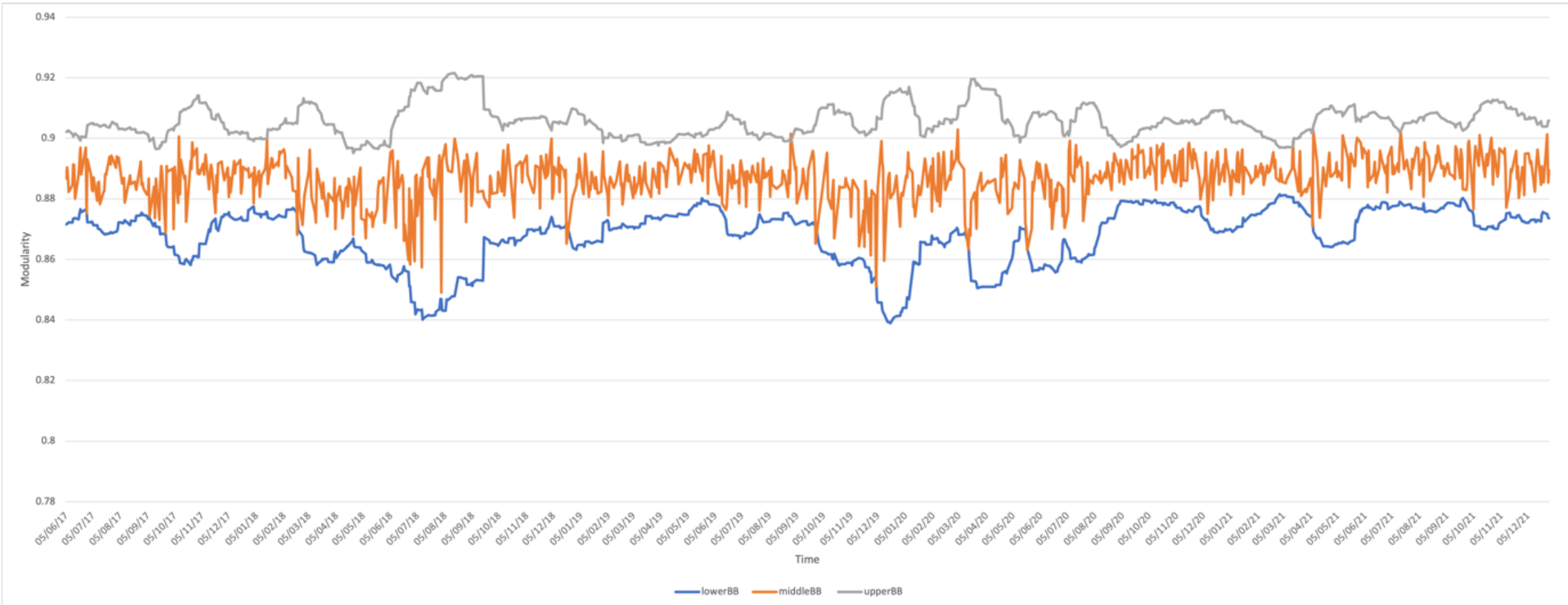

**Fig. 6** Bollinger Bands on the time series of hyperbolic modularity**;** The orange curve is middleBB denotes the 20-day SMA series; gray curve and the blue curve are upperBB and lowerBB, respectively denoting 3 standard deviations above and below the SMA series.

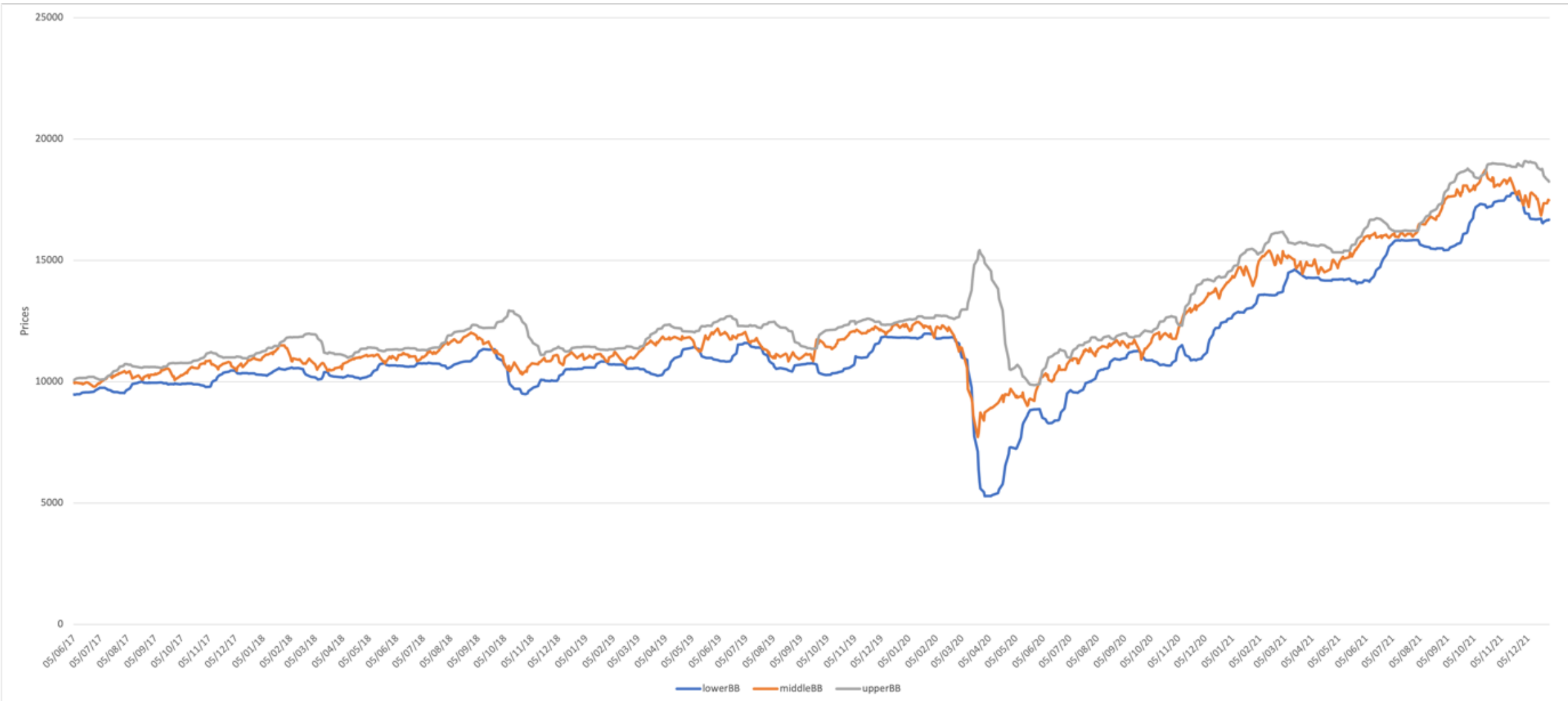

**Fig. 7** Bollinger Bands on the time series of CNX100 prices**;** The orange curve is middleBB denotes the 20-day SMA series; gray curve and the blue curve are upperBB and lowerBB, respectively, denoting 3 standard deviations above and below the SMA series.

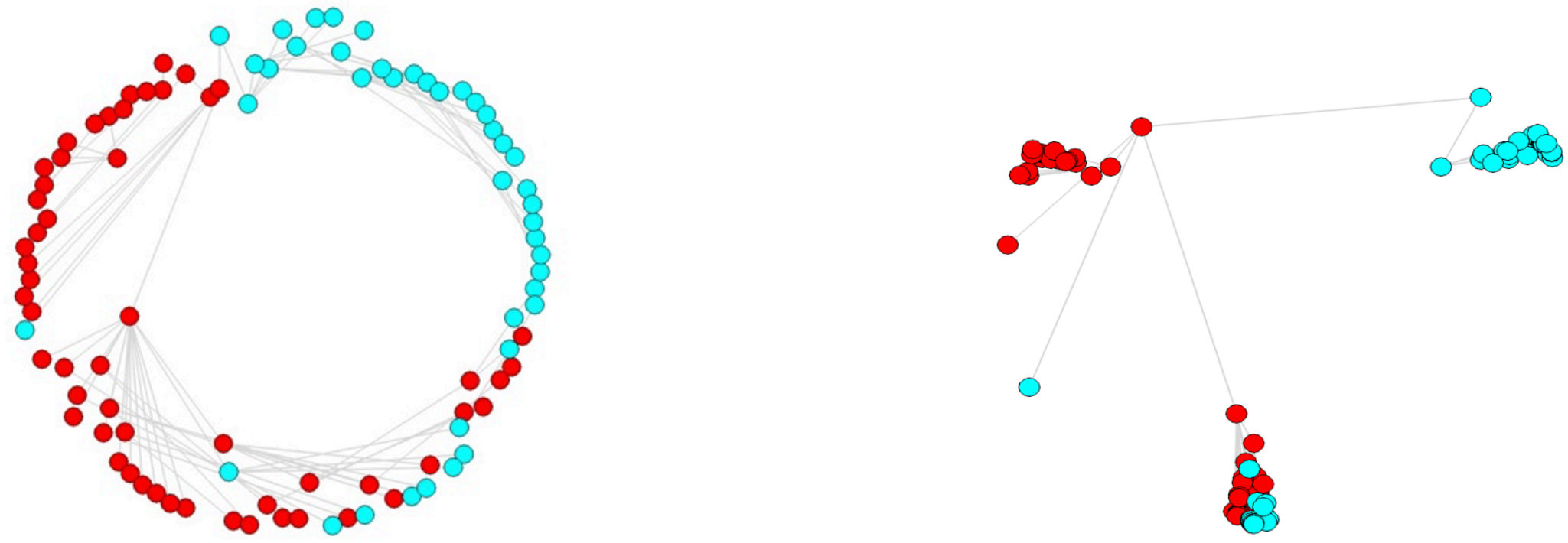

(a) (b)

**Fig. 8** Plot of two sectors; Finance (Red) and Healthcare (cyan) in the Poincaré disc via (a) ISO-EA (b) ISO-CA class of the coalescent embedding. Fig. 8(a) is showing a clear segregation.

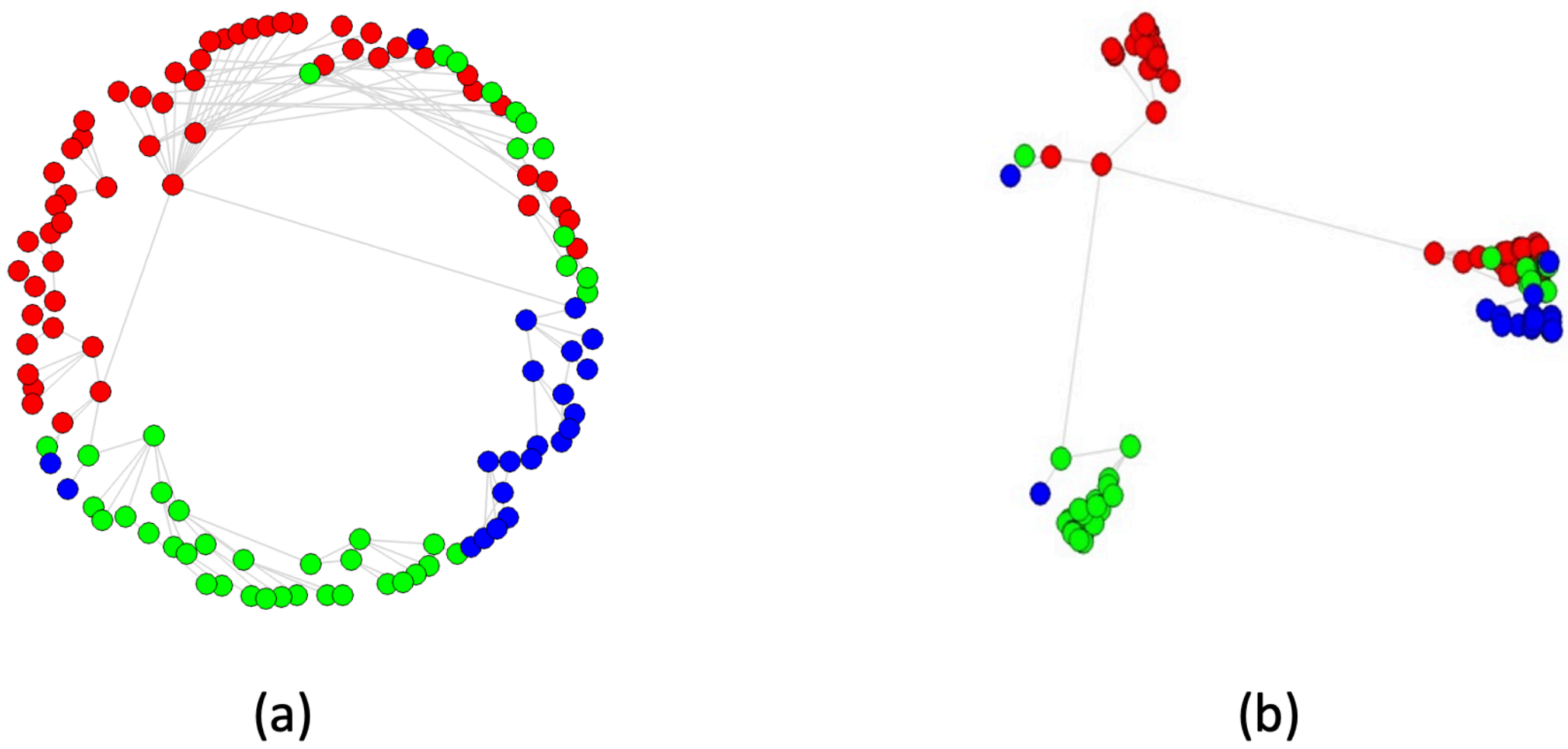

**Fig. 9** Plot of three sectors; Finance (Red), Healthcare (green) and Information technology (blue) in the Poincaré disc via (a) ISO-EA (b) ISO-CA class of the coalescent embedding. Fig. 9(a) is showing a clear segregation.

# Supporting Information

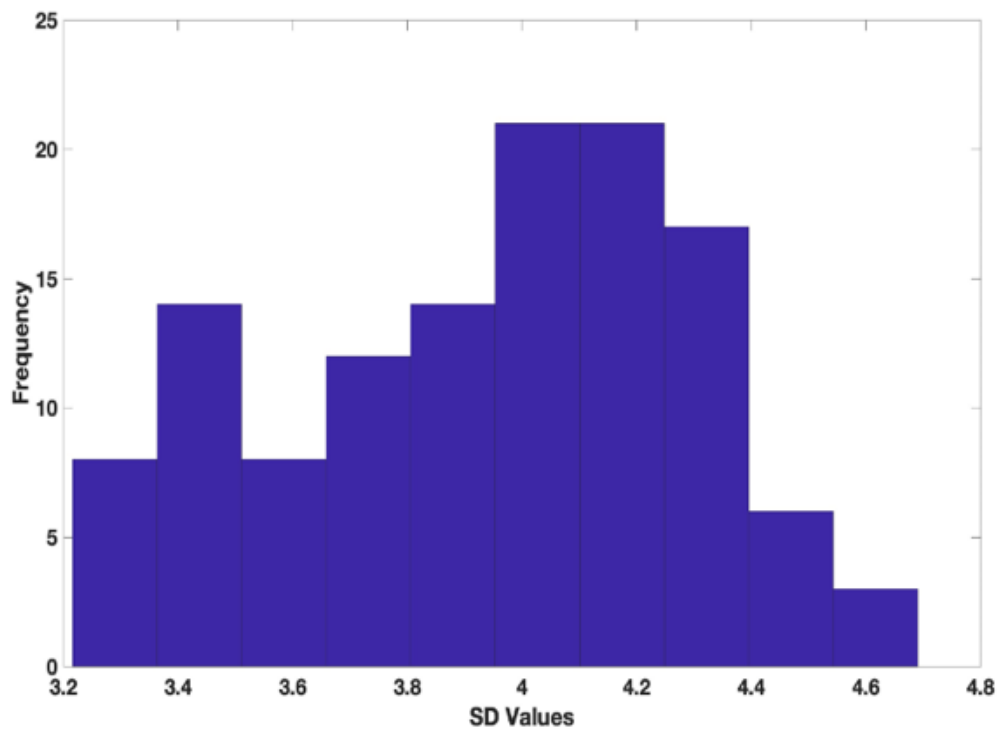

a) SD

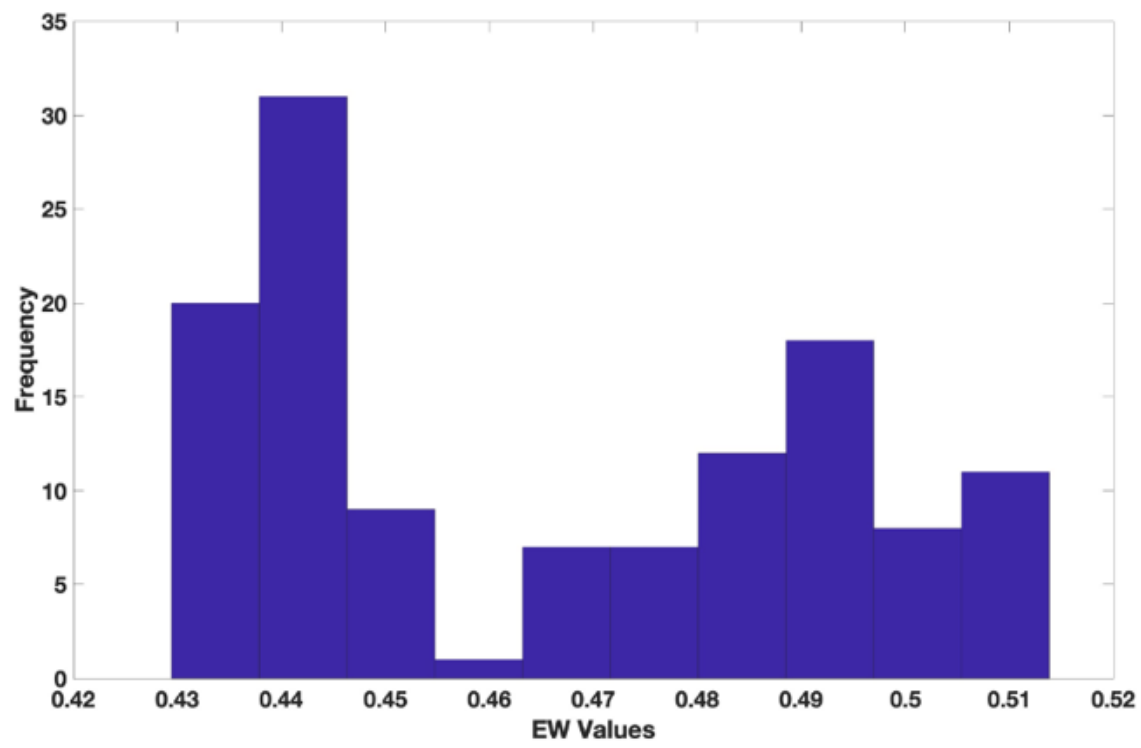

b) EW

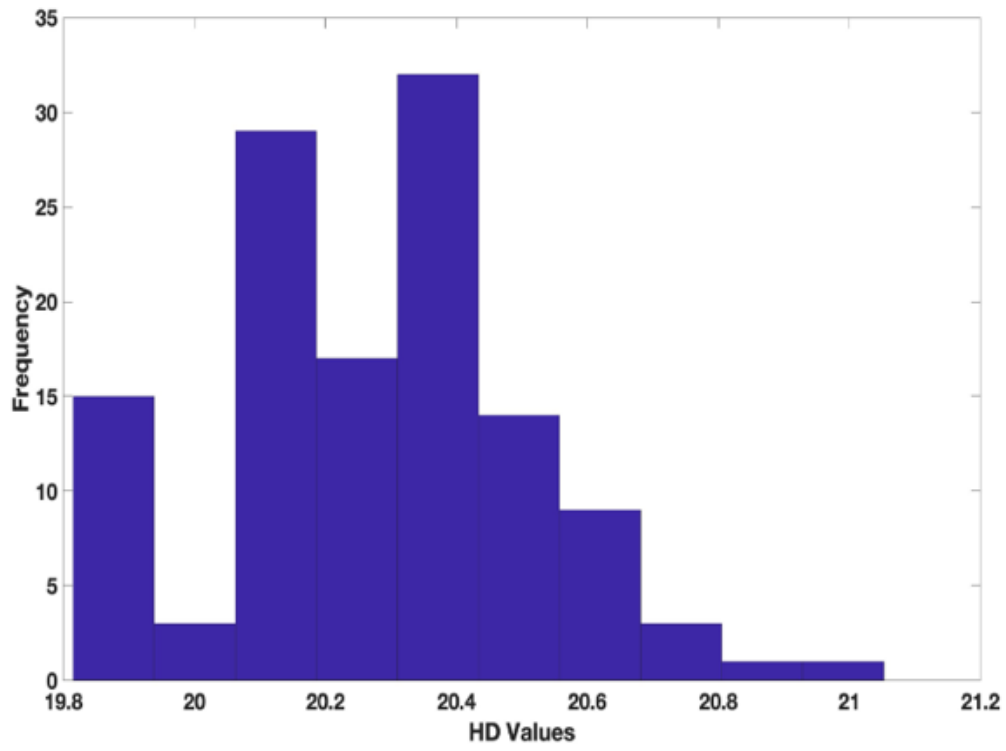

a) HD

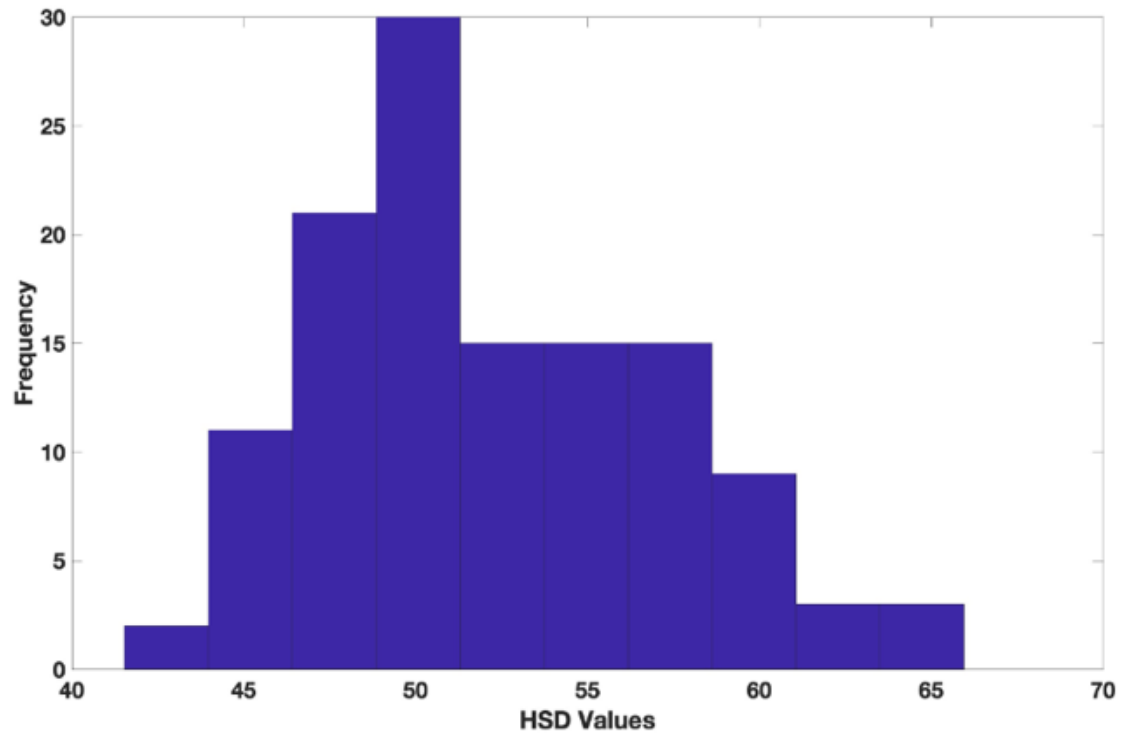

b) HSD

**S1 Fig** Probability distribution of the mean measures a) SD, b) EW of the original PMFG networks and mean measures c) HD, d) HSD of embedded PMFG networks (ISO-EA) for healthy period year 2018. The figures are suggestive that the various measures do not obey a power law distribution.

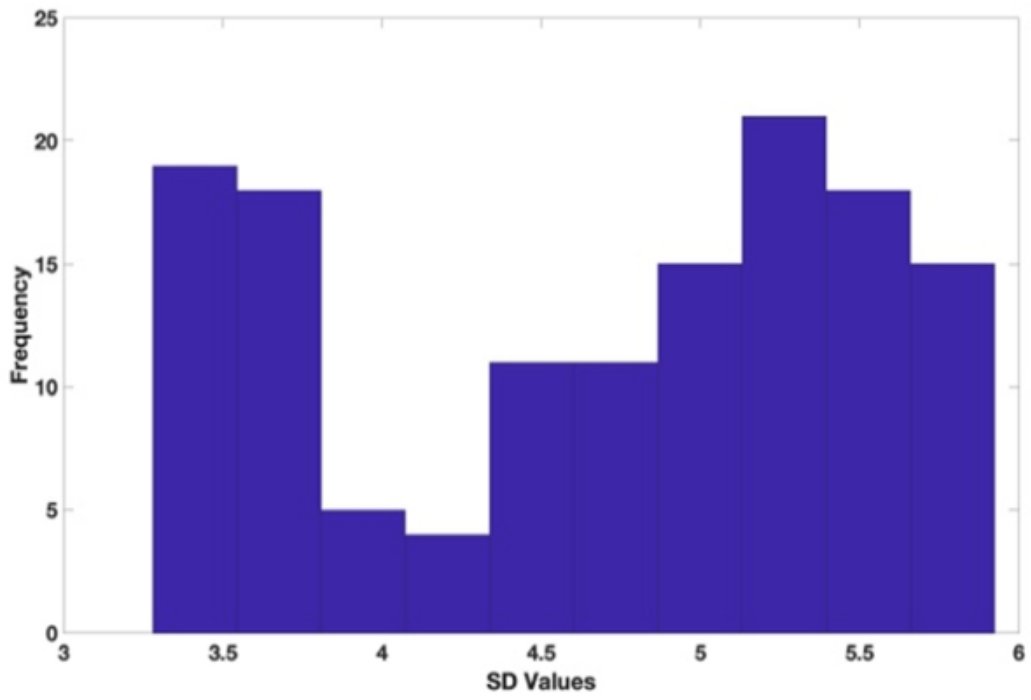

a) SD

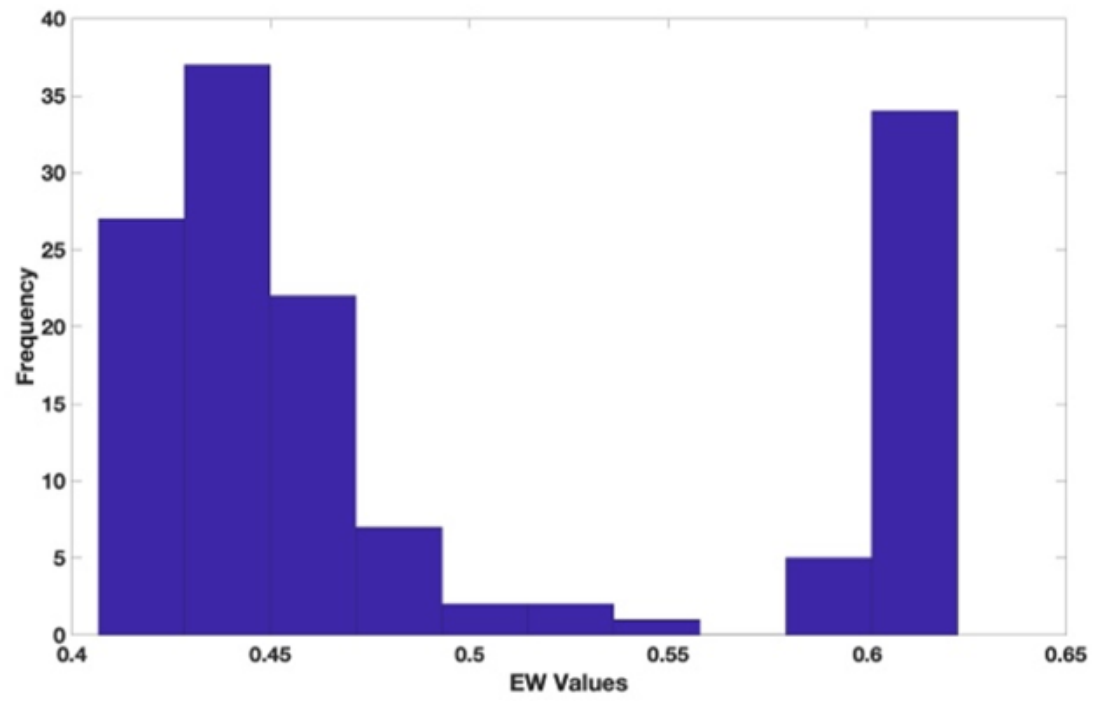

b) EW

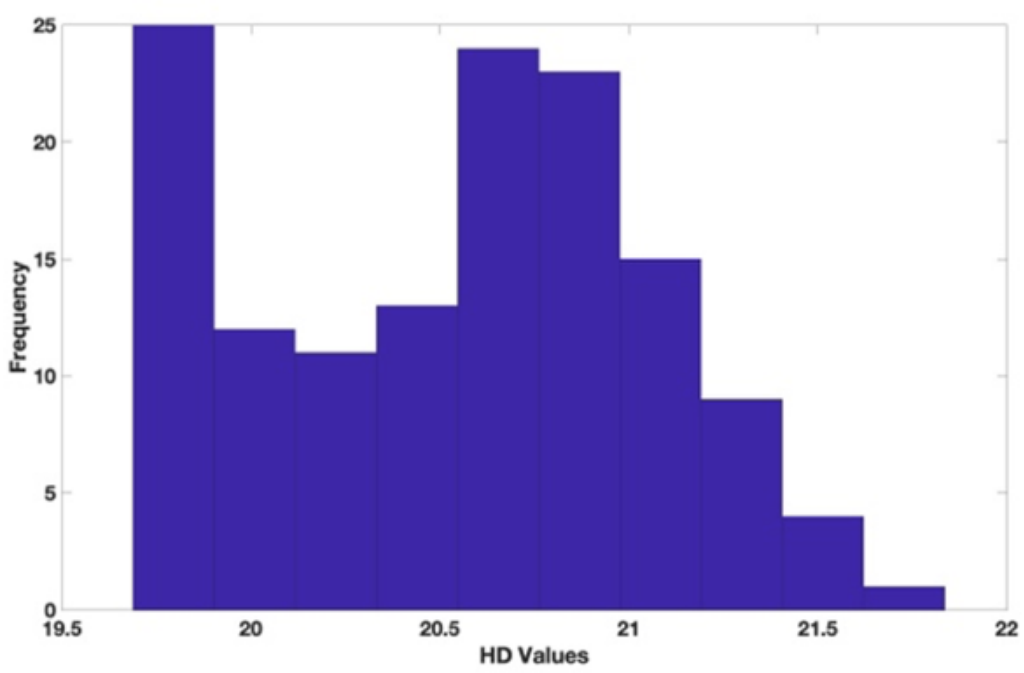

a) HD

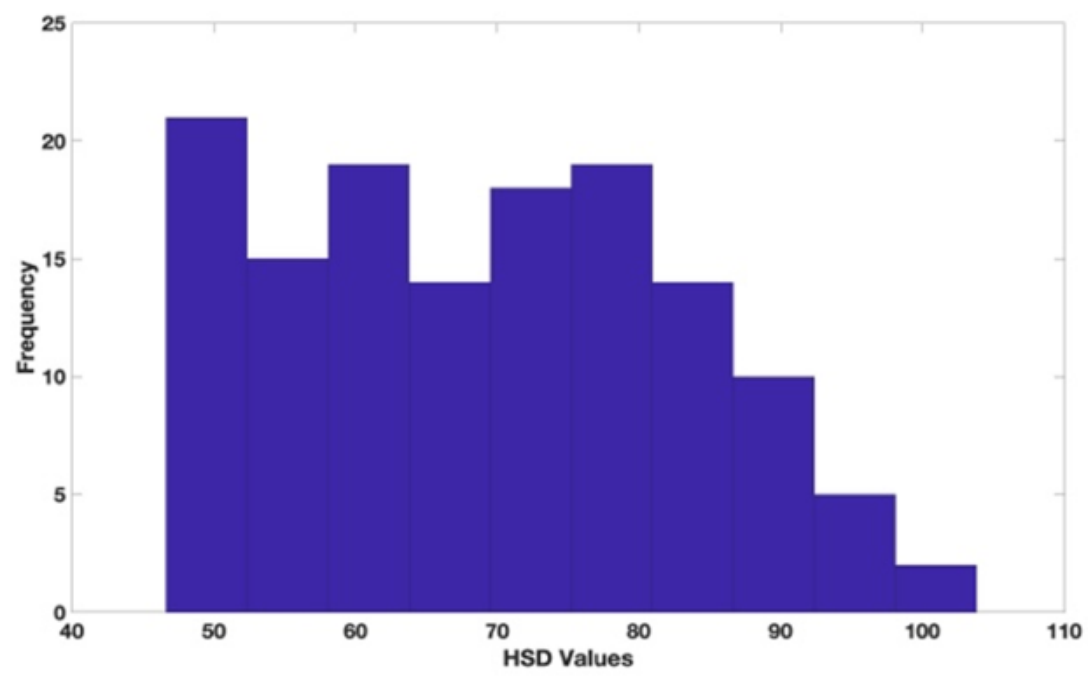

b) HSD

**S2 Fig** Probability distribution of the mean measures a) SD, b) EW of the original PMFG networks and mean measures c) HD, d) HSD of embedded PMFG networks (ISO-EA) for healthy period year 2020. The figures are suggestive that the various measures do not obey a power law distribution.

**S1 Table KS test** estimated power-law exponent $\gamma$ for degree distribution of Networks

| Year | MST - Power-law coefficient $\gamma$ ($p$-value) | PMFG - Power-law coefficient $\gamma$ ($p$-value) | PTN - Power-law coefficient $\gamma$ ($p$-value) |
|---|---|---|---|
| 2017 | 2.706 (0.355) | 2.575 (0.931) | 2.27400 (0.555) |
| 2018 | 2.687 (0.999) | 2.657 (0.699) | 2.3270(1.78E-35) |
| 2019 | 2.718 (0.449) | 2.613 (0.990) | 2.3020(0.00E+00) |
| 2020 | 2.638 (0.111) | 2.576 (0.994) | 2.294000(0.897) |
| 2021 | 2.880 (0.164) | 2.505 (0.176) | 2.2250(8.58E-33) |

The S1 Table presents the power-law coefficient along with KS test p-values for the networks MST, PMFG and PTN yearly. It is quite evident that p-values are greater than 10% in the case of MST and PMFG networks. Thus, MST and PMFG follows the power-law distribution but PTN.

**S2 Table** NMI Scores $NMI(C^{top}, C^{Euc})$ and $NMI(C^{top}, C^{hyp})$ for PMFG Networks

| Methods | 2017 | 2018 | 2019 | 2020 | 2021 |
|---|---|---|---|---|---|
| ISO | 0.6004 | 0.5531 | 0.7299 | 0.6601 | 0.6852 |
| ncISO | 0.4507 | 0.4395 | 0.5870 | 0.5085 | 0.4842 |
| LE | 0.0701 | 0.0739 | 0.0481 | 0.0861 | 0.0607 |
| MCE | 0.6128 | 0.5363 | 0.6198 | 0.6509 | 0.7014 |
| ncMCE | 0.5445 | 0.4156 | 0.6256 | 0.5953 | 0.6309 |
| Coal-ISO-EA | 0.5750 | 0.5482 | 0.6857 | 0.6253 | 0.6268 |
| Coal-ncISO-EA | 0.5940 | 0.5723 | 0.5927 | 0.6263 | 0.6336 |
| Coal-LE-EA | **0.6806** | **0.6070** | **0.6184** | **0.6661** | **0.7504** |
| Coal-MCE-EA | 0.4849 | 0.4438 | 0.5144 | 0.5405 | 0.5765 |
| Coal-ncMCE-EA | 0.5226 | 0.5313 | 0.5674 | 0.6765 | 0.6808 |

The rows 2 to 6 of the S2 Table gives the NMI score between the $C^{top}$ and $C^{Euc}$, and next rows 7 to 11 gives the NMI score between the $C^{top}$ and $C^{hyp}$. EA represents the equidistance adjustment. Note that LE-EA class of coalescent embedding is clear winner across all years, thus matches the hyperbolic cluster more closely then Euclidean clusters.

**S3 Table** AMI Scores $AMI(C^{top}, C^{Euc})$ and $AMI(C^{top}, C^{hyp})$ for PMFG Networks

| Methods | 2017 | 2018 | 2019 | 2020 | 2021 |
|---|---|---|---|---|---|
| ISO | 0.5819 | 0.5199 | 0.7239 | 0.6364 | 0.6713 |
| ncISO | 0.4252 | 0.3978 | 0.5786 | 0.4743 | 0.4609 |
| LE | 0.0186 | 0.0127 | 0.0168 | 0.0212 | 0.0091 |
| MCE | 0.5943 | 0.5026 | 0.6120 | 0.6267 | 0.6879 |
| ncMCE | 0.5231 | 0.3729 | 0.6173 | 0.5674 | 0.6145 |
| Coal-ISO-EA | 0.5565 | 0.5158 | 0.6794 | 0.5993 | 0.6099 |
| Coal-ncISO-EA | 0.5762 | 0.5411 | 0.5847 | 0.6022 | 0.6180 |
| Coal-LE-EA | **0.6666** | **0.5783** | **0.6110** | **0.6440** | **0.7397** |
| Coal-MCE-EA | 0.4620 | 0.4039 | 0.5049 | 0.5093 | 0.5584 |
| Coal-ncMCE-EA | 0.5018 | 0.4986 | 0.5591 | 0.6545 | 0.6673 |

The rows 2 to 6 of the S3 Table gives the AMI score between the $C^{top}$ and $C^{Euc}$, and next rows 7 to 11 gives the AMI score between the $C^{top}$ and $C^{hyp}$. EA represents the equidistance adjustment. Note that LE-EA class of coalescent embedding is clear winner across all years, thus matches the hyperbolic cluster more closely then Euclidean clusters.

**S4 Table** NMI dynamics of $NMI_{top}^{hyp}$ and $NMI_{top}^{Euc}$ for PMFG networks

| Methods | mean $NMI_{top}^{hyp}$ | mean $NMI_{top}^{Euc}$ | $p$-value |
|---|---|---|---|
| ISO, Coal-ISO-EA | 0.5562 | 0.6252 | 1 |
| ncISO, Coal-ncISO-EA | **0.5576** | **0.4947** | **5.00774697e-74** |
| LE, Coal-LE-EA | **0.6345** | **0.0635** | **7.86206154e-289** |
| MCE, Coal-MCE-EA | 0.4706 | 0.5992 | 1 |
| ncMCE, Coal-ncMCE-EA | **0.5217** | **0.4874** | **1.18882772e-18** |

The S4 Table, gives the mean of the NMI calculated over the 879 windows, comparing $C^{hyp}$ with $C^{top}$ ($NMI_{top}^{hyp}$) and NMI of $C^{Euc}$ with $C^{top}$ ($NMI_{top}^{Euc}$) for each class of the coalescent embedding algorithm. Additionally, table also presents the $p$-values obtained from the MW test, which are significant at the level of 0.05. Note, that the mean of $NMI_{top}^{hyp}$ is greater than the mean of $NMI_{top}^{Euc}$ across the three classes of the coalescent embedding.

**S5 Table** Healthy period versus Crisis period Mann Whitney Test $p$-values for PMFG networks

| Methods | Mean Measure Crisis Period (Year 2020) | Mean Measure Healthy Period (Year 2018) | $p$-value |
|---|---|---|---|
| ISO-CA-HD | 19.77301 | 19.72157 | 1.41E-01 |
| ISO-CA-HSD | 59.55195 | 46.08662 | 0.00E+00 |
| ISO-EA-HD | 20.55894 | 20.27156 | 6.45E-07 |
| ISO-EA-HSD | 69.18347 | 52.31449 | 0.00E+00 |
| ncISO-CA-HD | 19.90197 | 19.77596 | 6.60E-03 |
| ncISO-CA-HSD | 63.57935 | 48.49871 | 0.00E+00 |
| ncISO-EA-HD | 20.55894 | 20.27156 | 6.29E-07 |
| ncISO-EA-HSD | 70.44252 | 53.54735 | 0.00E+00 |
| LE-CA-HD | 19.61113 | 19.41034 | 2.45E-03 |
| LE-CA-HSD | 56.95834 | 43.65652 | 0.00E+00 |
| LE-EA-HD | 20.55894 | 20.27156 | 6.03E-07 |
| LE-EA-HSD | 68.84244 | 52.37834 | 0.00E+00 |
| MCE-CA-HD | 19.9156 | 19.38465 | 3.11E-10 |
| MCE-CA-HSD | 67.04745 | 50.91291 | 0.00E+00 |
| MCE-EA-HD | 20.55894 | 20.27156 | 4.08E-07 |
| MCE-EA-HSD | 71.5531 | 55.47197 | 0.00E+00 |
| ncMCE-CA-HD | 19.69202 | 19.26732 | 3.02E-08 |
| ncMCE-CA-HSD | 63.81771 | 48.26688 | 0.00E+00 |
| ncMCE-EA-HD | 20.55894 | 20.27156 | 4.55E-07 |
| ncMCE-EA-HSD | 70.8716 | 54.8257 | 0.00E+00 |
| **Average (EW)** | **0.4907** | **0.4657** | **0.33622** |
| **Average SD** | **4.6804** | **3.9469** | **4.1072e-12** |

The S5 Table, gives the mean values of Hyperbolic Distance (HD) and Hyperbolic Shortest Path Distance (HSD) of the embedded MST network constructed for both crisis and healthy periods, categorized by each class of coalescent embedding. Additionally, the table includes the mean values of Edge Weight (EW) and Shortest Path Distance (SD) for the original MST network during crisis and healthy periods. In this context, Equidistance Adjustment (EA) and Circular Adjustment (CA) are used. HD and HSD measure distances within the hyperbolic space, whereas EW and SD refer to the original network's edge weight and shortest path distance, respectively. The last column of the table reports the $p$-values obtained from the MW test for both the periods and corresponding to each class.

**S6 Table** ASI Score for topological communities of PMFG networks

| Methods | 2017 | 2018 | 2019 | 2020 | 2021 |
|---|---|---|---|---|---|
| ISO-EA | 0.745534 | 0.69326 | 0.72188 | 0.804965 | 0.777182 |
| ncISO-EA | 0.756806 | 0.733355 | 0.730274 | 0.807613 | 0.820136 |
| LE-EA | **0.855997** | **0.791137** | **0.814732** | **0.881431** | **0.879758** |
| MCE-EA | 0.557049 | 0.537714 | 0.629692 | 0.746074 | 0.736861 |
| ncMCE-EA | 0.617045 | 0.63987 | 0.626749 | 0.822581 | 0.803692 |

The S6 Table, gives the yearly ASI score for the topological communities corresponding to each class of the coalescent embedding algorithm. EA represent the equidistance adjustment. Clearly the LE-EA class is the clear winner.